\documentclass[preprints,article,accept,moreauthors,pdftex]{Definitions/mdpi} 

\firstpage{1} 
\pubvolume{7}
\issuenum{3}
\articlenumber{53}
\pubyear{2021}
\copyrightyear{2021}
\history{Received: 16 January 2021; Accepted: 23 February 2021; Published: 3 March 2021}
 
\usepackage{amssymb}
\usepackage{ulem}

\newcommand{\rem}[1] {\textcolor{red}{\sout{}}}
\newcommand{\add}[1] {\textcolor{black}{#1}}
\newcommand{\rep}[2] {{\textcolor{red}{\sout{}}}{\textcolor{black}{#2}}}

\Title{Scope Out Multiband Gravitational-Wave Observations of GW190521-Like Binary Black Holes with Space Gravitational Wave Antenna B-DECIGO}

\Author{
Hiroyuki Nakano $^{1}$
\orcidA{} 
Ryuichi Fujita $^{2,3}$
\orcidB{}
Soichiro Isoyama $^{4}$
\orcidC{}
Norichika Sago $^{5,6}$
\orcidD{}
}

\AuthorNames{Hiroyuki Nakano}
\AuthorNames{Soichiro Isoyama}
\AuthorNames{Ryuichi Fujita}
\AuthorNames{Norichika Sago}

\address{%
$^{1}$ \quad 
Faculty of Law, Ryukoku University, Kyoto 612-8577, Japan; \\ 
~~~\quad hinakano@law.ryukoku.ac.jp\\
$^{2}$ \quad 
Institute of Liberal Arts, Otemon Gakuin University, Osaka567-8502, Japan; \\
$^{3}$ \quad 
Center for Gravitational Physics, Yukawa Institute for Theoretical Physics, 
Kyoto University,\\
~~~\quad Kyoto 606-8502, Japan;\\
~~~\quad
ryuichi.fujita@yukawa.kyoto-u.ac.jp\\
$^{4}$ \quad 
School of Mathematics, University of Southampton, Southampton SO17 1BJ, United Kingdom; isoyama@yukawa.kyoto-u.ac.jp\\
$^{5}$ \quad 
Department of Physics, Kyoto University, Kyoto 606-8502, Japan; \\
$^{6}$ \quad 
Advanced Mathematical Institute, Osaka City University, Osaka 558-8585, Japan; sago@tap.scphys.kyoto-u.ac.jp\\
}

\corres{Correspondence: hinakano@law.ryukoku.ac.jp}

\abstract{
The gravitational wave event, GW190521 is the most massive binary black hole merger 
observed by ground-based gravitational wave observatories LIGO/Virgo to date. 
While the observed gravitational-wave signal is mainly in the merger and ringdown phases, 
the inspiral gravitational-wave signal of GW190521-like binary will be more visible 
by space-based detectors in the low-frequency band.
In addition, the ringdown gravitational-wave signal will be more loud with the next generation (3G)
of ground-based detectors in the high-frequency band, 
displaying a great potential of the multiband gravitational wave observations.
In this paper, we explore the scientific potential of multiband observations 
of GW190521-like binaries with milli-Hz gravitational wave observatory: LISA, 
deci-Hz observatory: B-DECIGO, and (next generation of) hecto-Hz observatories: aLIGO and ET.  
In the case of quasicircular evolution, 
the triple-band observation by LISA, B-DECIGO and ET will 
provide parameter estimation errors of the masses and spin amplitudes
of component black holes at the level of order 1\% -- 10\%. 
This would allow consistency tests of general relativity in the strong-field 
at an unparalleled precision, particularly 
with the ``B-DECIGO + ET'' observation.  
\rem{It would also enable to probe the ergoregion of the remnant Kerr BH through the measurement of quasinormal modes by ET.}
In the case of eccentric evolution, the multiband signal-to-noise ratio  
by ``B-DECIGO + ET'' observation would be larger than 100 for a five year observation prior to coalescence, 
even with high final eccentricities.
}


\begin{document}

\section{Introduction}

Among gravitational-wave (GW) events detected 
by LIGO and Virgo during O1, O2 and O3a runs~\cite{LIGOScientific:2018mvr,LIGOScientific:2020ibl}, 
a binary black hole (BBH) merger: GW190521~\cite{Abbott:2020tfl,Abbott:2020mjq} 
is one of the most striking discoveries. 
GW190521 is the heaviest BBH merger ever observed, 
producing the remnant black hole (BH) with the mass of $142^{+28}_{-16}\,M_{\odot}$ 
\footnote{
We set $G=1=c$ with the useful conversion factor $1 M_{\odot} = 1.477 \; 
{\mathrm {km}} = 4.926 \times 10^{-6} \; {\mathrm{s}}$. 
We also assume a ``Planck'' flat cosmology 
(when it is needed) 
with the Hubble constant $H_0 = 67.7 \,{\mathrm {km\, s^{-1}\, Mpc^{-1}}}$,  
and density parameters $\Omega_M = 0.307$ and $\Omega_\Lambda = 0.694$~\cite{Ade:2015xua}.} 
that can be interpreted as an intermediate mass BH; 
the source parameters of GW190521 (and our notations) are summarized in Table~\ref{table:GW190521}.  
This measurement triggers the intense investigation of GW190521's unique source property.

\begin{table}[ht]
\caption{Summary of the source parameters of GW190521 as a quasicircular BBH merger, 
reported by the LIGO-Virgo collaboration~\cite{Abbott:2020tfl}
based on \rep{a}{the} BBH waveform model~\cite{Varma:2019csw}. 
The symmetric 90\% credible interval for each parameter is also quoted.  
Note that the parameters are written in our notation, 
and masses are given in the source's rest frame; 
multiply by $(1 + z)$ to convert to the observer frame.}
\label{table:GW190521}
\centering
\begin{tabular}{lcc}
\toprule
Parameter 
& Symbol &
\\
\midrule
Primary mass [$M_{\odot}$] & $m_2^{\rm r}$ 
& $85^{+21}_{-14}$ 
\\
Secondary mass [$M_{\odot}$] & $m_1^{\rm r}$ 
& $66^{+17}_{-18} $ 
\\
Primary spin magnitude & $|\vec{\chi}_2|$ 
& $0.69^{+0.27}_{-0.62}$ 
\\
Secondary spin magnitude & $|\vec{\chi}_1|$ 
& $0.73^{+0.24}_{-0.64}$ 
\\
Total mass [$M_{\odot}$] & $m_t^{\rm r}~(=m_1^{\rm r}+m_2^{\rm r})$ 
& $150^{+29}_{-17}$ 
\\ 
Mass ratio & $q~(= m_1^{\rm r}/m_2^{\rm r} \leq 1)$ 
& $0.79^{+0.19}_{-0.29}$ 
\\
Luminosity Distance [Gpc] & $D_L$ 
& $5.3^{+2.4}_{-2.6}$ 
\\
Redshift & $z$ 
& $0.82^{+0.28}_{-0.34}$ 
\\
\bottomrule
\end{tabular}
\end{table}

A key element to better understand GW190521 is the precise measurement of the binary parameters
(see, for example, Ref.~\cite{Nitz:2020mga} for a possibility of
an intermediate mass ratio inspiral).  
GW190521 is, however, \add{a} much heavier BBH system than previously observed GW events, 
and one of the difficulties here is the short duration and bandwidth 
of the GW signal that can be observed in the LIGO/Virgo band.  
In the case of the quasicircular BBH scenario 
(which is most favored by the LIGO/Virgo analysis~\cite{Abbott:2020tfl,Abbott:2020mjq}), 
the coalescing time and the \rep{numbers of cycle at the GW frequency}{number of GW cycles at frequency} $f$ (in the observer frame) 
are estimated as 
\begin{align}
 \label{eq:tc}
 t_c &\sim  1.3 \, (1+z)^{-5/3} 
 \left(\frac{m_1^{\rm r}}{66M_\odot}\right)^{-1}
 \left(\frac{m_2^{\rm r}}{86M_\odot}\right)^{-1} \left(\frac{m_t^{\rm r}}{150M_\odot}\right)^{1/3}
 \left(\frac{f}{10.0\,{\rm Hz}}\right)^{-8/3} \,{\rm s}\,, \\
 N_c &\sim 1.0 \times 10^{1} \, (1+z)^{-5/3} 
 \left(\frac{m_1^{\rm r}}{66M_\odot}\right)^{-1}
 \left(\frac{m_2^{\rm r}}{85M_\odot}\right)^{-1} \left(\frac{m_t^{\rm r}}{150M_\odot}\right)^{1/3}
 \left(\frac{f}{10.0\,{\rm Hz}}\right)^{-5/3} \,, 
 \label{eq:Nc}
\end{align}
indicating the lack of the GW signals from the sufficiently long inspiral phase. 
Because of the short duration signal dominated 
by the merger and ringdown phases, for example, only weak constraints are obtained 
for the component BH spins and their orientations~\cite{Abbott:2020mjq}. 
Furthermore, even alternative interpretations of the observed GW signal 
other than massive quasicircular BBH merger in general relativity (GR) 
would become more relevant; several plausible scenarios are assessed 
in Section~6 of Ref.~\cite{Abbott:2020mjq} by the LIGO-Virgo collaboration.

\begin{figure}[ht]
\centering
\includegraphics[width=0.85\textwidth]{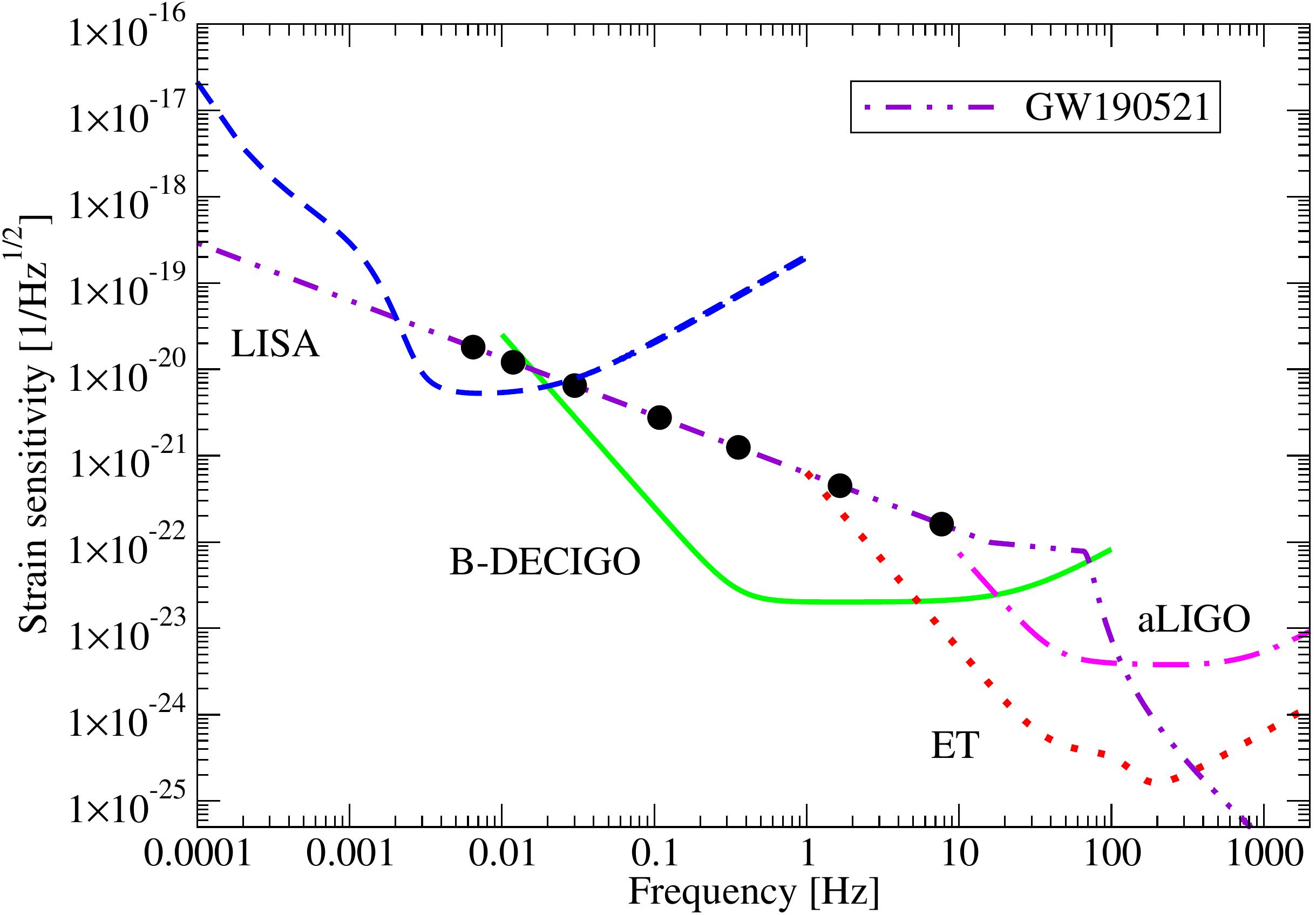}
\caption{Strain sensitivity curves for \add{the} ground-based aLIGO
and a next-generation (3G) detector: 
Einstein Telescope (ET), 
and space-based B-DECIGO and LISA,
together with the GW amplitude of \add{a} GW190521-like \add{nonprecessing}, quasicircular BBH. 
We use the median values in Table~\ref{table:GW190521} as source parameters.
The noise power spectral density (PSD) for each GW detector 
is given in Section~\ref{subsec:detectors}, 
and the spectral density of \add{the} BBH amplitude is obtained 
by the nonspinning, IMR amplitude model in Section~\ref{subsec:SNR}. 
As a reference, we mark \add{with} the black dots with 5 years, 1 year, 1 month, 1 day, 1 hour, 1 minute 
and 1 second before merger time~\eqref{eq:tc}, from the left to the right, respectively.}
\label{fig:N_curve}
\end{figure}

At the same time, the estimation in Eqs.~\eqref{eq:tc} and~\eqref{eq:Nc} 
suggests a natural way to overcome the hurdle here: 
observe the inspiral GW signal in the low-frequency band 
offered by space-based GW detectors. 
Future GW astronomy in 2030s will utilize the LISA observatory 
in the milli-Hz band~\cite{Audley:2017drz} 
and deci-Hz GW detectors such as B-DECIGO~\cite{Nakamura:2016hna}:  
a prototype GW antenna of the DECIGO mission~\cite{Seto:2001qf,Kawamura:2020pcg}
\footnote{
Other proposed GW missions in the low-frequency band, 
including Taiji~\cite{Guo:2018npi} and TianQin~\cite{Luo:2015ght} 
in the milli-Hz band, 
and MAGIS~\cite{Graham:2017lmg} and TianGO~\cite{Kuns:2019upi} 
in the deci-Hz band are concisely summarized in, 
for example, 
reviews by Ni~\cite{Ni:2016wcv,Ni:2020utm}.
}.
Figure~\ref{fig:N_curve} plots the track of the strain sensitivity curve 
of GW190521-like non-spinning BBH system, assuming the quasicircular evolution
and a simple inspiral--merger--ringdown (IMR) amplitude model given in Eq.~\eqref{eq:IMR}. 
At the GW frequency $f = 0.1$ Hz, for example, 
Eqs.~\eqref{eq:tc} and~\eqref{eq:Nc} give 
the \rep{coalescing}{coalescence} time 
$t_c \sim 1.1 \times 10^5\,{\rm s}$ and 
the \rep{numbers of cycles}{number of GW cycles} $N_c \sim 8.1 \times 10^3$ 
with the dimensionless characteristic strain 
\begin{align}
\label{eq:hc}
  h_c \sim 7.6\times 10^{-22}\,(1+z)^{5/6} \left(\frac{{\cal M}^{\rm r}}{65.1M_\odot}\right)^{5/6}
 \left(\frac{f}{0.1\,{\rm Hz}}\right)^{-1/6} \left(\frac{D_L}{5.3\,{\rm Gpc}}\right)^{-1} \,,
\end{align}
where we introduce the \rem{the} source's rest-frame chirp mass 
${\cal M}^{\rm r} \equiv \nu^{3/5}\,m_t^{\rm r}$ 
with the symmetric mass ratio $\nu \equiv  {m_1^{\rm r} m_2^{\rm r}}/{{m_t^{\rm r}}^2}$. 
In fact, we find in Section~\ref{sec:params-err} that 
the sky and polarization averaged signal-to-noise ratio 
(SNR) 
(whose meaning is momentarily clarified in Section~\ref{sec:method}) 
accumulated $5$-years before the final coalescence would be $\sim 5.9 \times 10^{1}$ 
in the B-DECIGO band and $\sim 2.7$ in the LISA band. 
These estimations show that the early-inspiral signal of GW190521-like BBHs 
would be sensitive in the LISA band, and it would be even loud in the deci-Hz band.

Although the observation with LISA and B-DECIGO alone 
would provide \rem{a} valuable information of the inspiral 
GW signal from the GW190521-like BBH system, 
the true potential of \rep{their}{having the} low-frequency sensitivity will be revealed only when it is combined with the high-frequency sensitivity in the hecto-Hz band. 
As illustrated in Figure~\ref{fig:N_curve}, 
the late-inspiral and merger--ringdown GW signals 
of GW190521-like BBH are best detected with aLIGO, 
Virgo and KAGRA~\cite{Akutsu:2020his}.
In addition to ground-based GW observatories
that are online,  
the next generation (3G) of ground-based detectors such as 
Einstein Telescope (ET)~\cite{Hild:2009ns,Punturo:2010zz} 
(see also Ref.~\cite{Reitze:2019iox}
for Cosmic Explorer (CE)) 
will significantly improve the visibility of GW190521-like BBH systems. 
We will see in Section~\ref{sec:params-err} that 
the averaged SNR of the late-inspiral and ringdown signals 
in the ET band 
would be $\sim 2.7 \times 10$ and $\sim 1.5 \times 10^2$, respectively. 
The joint ``space $+$ ground'' observation across the full GW bands
would be therefore the best way to observe the GW190521-like BBH systems; 
this is the basic idea of \textit{multiband GW astronomy}.

Soon after the first detection of GW150914~\cite{Abbott:2016blz}, 
the potential of multiband GW astronomy of BBH systems 
with LISA and aLIGO \rep{is}{was} emphasized~\cite{Sesana:2016ljz}
(see also Ref.~\cite{Cutler:2019krq}). 
This study \rem{is}{was} immediately followed up with more detailed analyses. 
The works include (but not limited to) 
the improved estimation of source parameters~\cite{Vitale:2016rfr,Jani:2019ffg}, 
tests of GR with high precision~\cite{Barausse:2016eii,Carson:2019rda,Carson:2019kkh,Datta:2020vcj,Gupta:2020lxa,Toubiana:2020vtf}, refined event rate estimations~\cite{Gerosa:2019dbe}, 
probing environment effects~\cite{Ng:2020jqd,Toubiana:2020drf}, 
and new data analysis ideas~\cite{Ewing:2020brd}; 
all prove the scientific values added by the multiband observation. 
Among these investigations, Refs.~\cite{Nair:2015bga,Nakamura:2016hna,Nair:2018bxj,Isoyama:2018rjb,Grimm:2020ivq,Liu:2020nwz} 
demonstrated that the multiband observation of stellar-mass BBH systems 
will further benefit from having (B-)DECIGO in the deci-Hz band, 
which naturally bridges the gap between LISA and aLIGO bands.

\subsection{Goals and organization of this paper}

Our purpose in this paper is to explore the prospects for the multiband observation 
of GW190521-like \rep{non-precessing}{nonprecessing}, quasicircular ``intermediate-mass'' BBHs. 
The possibility of \rem{the} multiband observations of intermediate-mass BBHs 
with deci-Hz GW detectors was pointed out 
by Amaro-Seoane and Santamaria~\cite{AmaroSeoane:2009ui},
and Yagi~\cite{Yagi:2012gb}. 
We consider these observations across the full GW spectrum 
provided by LISA in the milli-Hz band, B-DECIGO in the deci-Hz band, 
and aLIGO and ET in the hecto-Hz, 
looking at two specific aspects of \rem{the} multiband GW astronomy/physics: 
parameter estimation errors and tests of GR.

We begin in Section~\ref{sec:method} 
by providing a set of our basic tools for the signal analysis
in the matched filtering technique:
i) the noise PSD in Section~\ref{subsec:detectors}; 
ii) the GW signal models in Sections~\ref{subsec:GW} and~\ref{subsec:SNR}; 
and iii) the multiband Fisher matrix formalism in Section~\ref{subsec:Fisher}. 
In Section~\ref{sec:params-err}, we present the parameter estimation errors 
of GW190521-like \rep{non-precessing}{nonprecessing}, quasicircular BBH systems using \rem{the} multiband GW observations. 
They are displayed in Table~\ref{table:sta_err_GW190521} and~\ref{table:sta_err_RD}.
Based on the estimated errors, we then examine in Section~\ref{sec:implication} 
\add{to} what extent the future multiband \rep{observation}{observations} of GW190521-like BBH \rep{system}{systems} 
will improve tests of GR; this includes an 
inspiral--ringdown consistency test (see Table~\ref{table:IMR_test}), 
and a simple test to discriminate between the remnant BH based on GR 
and other remnant compact objects (see Figure~\ref{fig:simple_QNM})\rem{,  
which may allow to prove the ergoregion of Kerr geometry shown in Figure~\ref{fig:NNT1}}.
We conclude in Section~\ref{sec:discussion} with complications and various effects, 
which are not covered in this paper due to our
assumptions and simplifications.

Appendices contain some additional analysis and information.
While \rep{the}{a} quasicircular BBH merger is the most favored scenario 
by the LIGO/Virgo analysis~\cite{Abbott:2020tfl, Abbott:2020mjq}, 
other alternative \rep{scenario}{scenarios} such as an eccentric BBH merger may also be consistent 
with the observed source for GW190521~\cite{Gayathri:2020coq}. 
We briefly discuss the multiband visibility of \add{a} GW190521-like eccentric BBH 
in Appendix~\ref{app:ecc}. 
Also, additional noise \rep{PSD}{PSDs} of ground/space-based, current/future GW detectors \rep{is}{are} provided 
in Appendix~\ref{app:various_sensitivity}, 
as a complement of our treatment in Section~\ref{subsec:detectors}.

Throughout this paper, the binary parameters and GW frequencies measured 
in the source's rest frame are denoted with the index `r', explicitly 
distinguished from those in the observer frame.

\section{Method for signal-to-noise ratio and Fisher analysis}
\label{sec:method}

In this section, we summarize our methodology of multiband GW data analysis  
in the specific context of a \rep{non-precessing}{nonprecessing}, spinning, quasicircular GW190521-like BBH system, 
where the spins are aligned to the orbital angular momentum; 
an alternative scenario\rem{s} of non-zero orbital eccentricity in 
GW190521-like BBH system will be discussed in Appendix~\ref{app:ecc}.
Our simple framework here is largely based on the previous 
B-DECIGO works~\cite{Nakamura:2016hna,Isoyama:2018rjb}.

\subsection{aLIGO, ET, B-DECIGO and LISA}
\label{subsec:detectors}

For our multiband GW observation, we follow Ref.~\cite{Isoyama:2018rjb} 
by considering the four GW observatories: 
aLIGO and ET in the hecto-Hz band, B-DECIGO in the deci-Hz, 
and LISA in the milli-Hz band 
(see also Appendix~\ref{app:various_sensitivity} for some of other GW detectors in these bands). 
It should be noted that we shall \rep{treat}{use} the ``non sky-\rep{averaging}{averaged}'' PSD $S_n(f)$; 
we will account for the average over the GW detector's antenna pattern function 
at the level of the waveform.

For aLIGO in the hecto-Hz band, we use Eq.~(4.7) in Ref.~\cite{Ajith:2011ec},
\begin{align}
S_n^{\rm aLIGO} =& 10^{-48} \left(
 0.0152\,{x}^{-4}+ 0.2935\,{x}^{9/4}+ 2.7951\,{x}^{3/2}- 6.5080\,{x}^{3/4}+ 17.7622 
 \right)~~{\rm Hz}^{-1} \,;
\cr
 x=& \frac{f}{245.4\,{\rm Hz}} \,.
\end{align}
For a 3rd generation (3G) GW interferometer: ET in the hecto-Hz band,
we find Table 1 in Ref.~\cite{Sathyaprakash:2009xs} as
\begin{align}
\label{noise-ET?}
S_n^{\rm ET} =& 
1.5\times 10^{-52}
\biggl(
{y}^{- 4.1}+ 186.0\,{y}^{- 0.69} 
\cr & + 233.0 \times {\frac { 1.0+ 31.0\,y- 65.0\,
{y}^{2}+ 52.0\,{y}^{3}- 42.0\,{y}^{4}+ 10.0\,{y}^{5}+ 12.0\,{y}^{6}}{
 1.0+ 14.0\,y- 37.0\,{y}^{2}+ 19.0\,{y}^{3}+ 27.0\,{y}^{4}}}
\biggr)~~{\rm Hz}^{-1}  \,;
\cr
 y=& \frac{f}{200.0\,{\rm Hz}} \,.
\end{align}
For B-DECIGO in the deci-Hz band,
we use Eq.~(20) in Ref.~\cite{Isoyama:2018rjb},
originally proposed by Nakamura et al.~\cite{Nakamura:2016hna}:
\begin{align}
S_n^{\rm BD} =& 
\left[(2.01 \times 10^{-23})^2 
+ \left(\frac{2.53 \times 10^{-18}}{F^2}\right)^2 
+ \left(8.00 \times 10^{-22}\, F \right)^2\right]~~{\rm Hz}^{-1}  \,;
\cr
 F=& \frac{f}{1.0\times 10^{-3}\,{\rm Hz}} \,.
\end{align}
For LISA in the milli-Hz band, we use Eq.~(1) of in Ref.~\cite{Cornish:2018dyw}, 
which is based on the 2018 LISA Phase-0 reference design parameters. 
\rem{Discarding the galactic confusion noise for simplicity,} 
It reads 
\footnote{Our expression $S_n^{\rm LISA}$~\eqref{S-LISA} is smaller than 
Eq.~(1) in Ref.~\cite{Cornish:2018dyw} by a overall factor of $5$; 
the difference arises simply because Eq.~\eqref{S-LISA} does not account for 
the sky-averaging. 
}
\begin{align}
\label{S-LISA}
S_n^{\rm LISA} =&
\biggl[ \biggl( {2.4602\times 10^{-41}}
+{4.0504 \times 10^{-38}}\,{z}^{2}
+\frac{ 4.7850\times 10^{-48}}{{z}^{2}}
\cr & 
+\frac{ 2.8485\times 10^{-51}}{{z}^{4}}
+\frac{ 3.9412\times 10^{-58}}{{z}^{6}}\biggr) + S_c^{\rm LISA} \biggr]~~{\rm Hz}^{-1} \,;
\cr
z=& \frac{f}{1.0\,{\rm Hz}} \,.
\end{align}
\add{where $S_c^{\rm LISA}(f)$ represents the effective PSD 
due to the unresolved galactic binaries; the explicit expression is given 
in Eq.~(14) of Ref.~\cite{Cornish:2018dyw} 
and we have assumed a $4$-year mission for Figure~\ref{fig:N_curve}.}

\subsection{Waveform models}
\label{subsec:GW}

We employ as our BBH waveform model the frequency-domain, ``restricted'' waveform 
in the stationary phase approximation for the inspiral phase, 
and the frequency-domain, single-mode waveform for the ringdown phase. 
We shall restrict our waveform model to these two phases to simplify our analysis 
as far as possible; 
the complete IMR treatment at the level of waveform 
(using, for example, \add{``effective-one-body'' (EOB) approach~\cite{Nagar:2018zoe,Ossokine:2020kjp} 
and ``phenomenological'' (IMRPhenom) model~\cite{Khan:2019kot,Pratten:2020fqn})} will be left for future work.

The GW waveform from a BBH inspiral 
in the frequency domain has
the well-known form (see, for example, Ref.~\cite{Berti:2004bd})  
\begin{equation}\label{eq:hf}
{\tilde h}_{\rm Insp}(f) = {\cal A} f^{-7/6} e^{i \Psi_{\mathrm {Insp}}(f) }\,, 
\end{equation}
where ${\cal A}$ is the ``Newtonian'' amplitude \rep{averaging}{averaged} 
over all sky positions and binary orientations 
(see, for example, Ref.~\cite{Dalal:2006qt}), so that 
\begin{equation}
\label{def-A}
{\cal A} \equiv
\frac{2}{5}  \sqrt{\frac{5}{24}} \pi^{-2/3} 
\frac{{\cal M}^{5/6}}{D_{L}}\,. 
\end{equation}
The waveform's frequency-domain phase $\Psi_{\mathrm {Insp}}(f)$ in the post-Newtonian (PN) approximation
is given by 
(see, for example, Ref.~\cite{Mishra:2016whh})
\begin{align}\label{eq:F2-BBH}
\Psi_{\mathrm {Insp}}(f)
&=   
2 \pi f t_c - \Psi_c - \frac{\pi}{4} 
+
\frac{3}{128 \nu v^5}
\left( 
\Delta \Psi^{\mathrm {pp}}_{3.5{\mathrm {PN}}}
+
\Delta \Psi^{\mathrm {pp-spin}}_{3.5{\mathrm {PN}}}
+
\Delta \Psi^{\mathrm {BH-tidal}}_{3.5{\mathrm {PN}}} 
\right)\,,
\end{align}
where $v \equiv ( \pi m_t f )^{1/3}$ is the PN parameter 
(in terms of the observer-frame total mass), 
and $t_c$ and \rep{$\phi_c$}{$\Psi_c$} are the time and phase at coalescence.
The phase terms $\Delta \Psi^{\mathrm {pp}}_{3.5{\mathrm {PN}}}$
and $\Delta \Psi^{\mathrm {pp-spin}}_{3.5{\mathrm {PN}}}$
are the $3.5$PN spin-independent, point-particle contributions derived in Ref.~\cite{Arun:2004hn}
and the $3.5$PN spin-dependent, point-particle contributions 
that include linear spin-orbit~\cite{Blanchet:2013haa,Bohe:2013cla}, 
quadratic-in-spin~\cite{Bohe:2015ana} 
and cubic-in-spin~\cite{Marsat:2014xea} effects, respectively.
The remaining phase \rep{tern}{term} $\Delta \Psi^{\mathrm {BH-tidal}}_{3.5{\mathrm {PN}}}$ 
is related to the tidal response of a spinning BH as a finite-size body, 
i.e., BH-absorption corrections 
such as the GW energy and angular momentum fluxes 
down to the horizons and the associated evolution 
of the BH itself~\cite{Chatziioannou:2016kem,Isoyama:2017tbp,Hughes:2018qxz}.

Meanwhile, limited to only fundamental $(n = 0)$, $\ell = 2 = m$ mode\rem{s}, 
the time-domain, \rep{signle}{single}-mode ringdown waveform measured at a GW observatory
is \add{written as~\cite{Nakano:2003ma,Nakano:2004ib,Berti:2005ys}}
\begin{align}
h_{\rm Ring}(f_c,\,Q,\,t_0,\,\phi_0;\,t) 
=
\begin{cases}
e^{ - \frac {\pi \,f_c\,(t-t_0)}{Q}}\,
\cos \left[2\,\pi \,f_c\,(t-t_0)-\phi_0 \right]
& {\rm for}~t \geq t_0 \,, \\ \\
0 & {\rm for}~t < t_0 \,,
\end{cases}
\label{eq:RDwave}
\end{align}
where $t_0$ and $\phi_0$ are the initial time and phase of the ringdown, respectively, 
and we have ignored the overall amplitude so that Eq.~\eqref{eq:RDwave} is not normalized; 
the initial ringdown amplitude may be determined by matching the ringdown GW waveform
to the merger one 
(refer to, for example, Eq.~\eqref{eq:IMR} below).
When the final remnant object is a Kerr BH, 
the central frequency $f_c$ and the quality factor $Q$ are given in terms of \rep{QNM}{quasinormal-mode (QNM)} frequencies
($f_{\rm QNM} = f_{\rm R} + i\, f_{\rm I}$) of the remnant BH as 
[see Eq.~(7) of Ref.~\cite{Nakano:2015uja}]
\begin{equation}
f_c := f_{\rm R}\,,
\quad
Q := -\frac{f_{\rm R}}{2 f_{\rm I}} \,.
\label{eq:f_Q}
\end{equation}
It should be noted that $f_{\rm QNM} = f_{\rm QNM}^{\rm r} (1 + z)^{-1}$ here 
is given in the observer frame. 
The Fourier transforms of Eq.~\eqref{eq:RDwave} provide the corresponding frequency-domain waveform, 
which takes the form (here we follow the convention of Ref.~\cite{Nakano:2003ma})
\begin{equation}
\label{eq:f-RDwave}
\tilde{h}_{\rm Ring}(f_c,\,Q,\,t_0,\,\phi_0;\,f) 
= {\frac {Q \left(  f_c\cos \phi_0 
-2\,i\, Qf\cos \phi_0 +2\,Q f_c\,\sin \phi_0 \right) }
{\pi\, \left( f_c+2\, i\,Q f_c-2\, i\,Qf \right)  
\left( f_c-2\, i\,Q f_c-2\,i\,Qf \right) }} 
\,{{\rm e}^{2\,i\pi\,f\,t_0}}\,.
\end{equation}
We note that this frequency-domain GW waveform is not normalized\rep{, too}{either} 
\footnote{The maximized SNR over the initial ringdown phase $\phi_0$ 
has been discussed in Ref.~\cite{Mohanty:1997eu}.},
following the approach presented in Appendix B of Ref.~\cite{Nakano:2004ib}. 


The ringdown waveforms~\eqref{eq:RDwave} and~\eqref{eq:f-RDwave} require
input data for the QNM frequency of the final remnant BH. 
Because we do not consider the complete IMR phase 
at the level of waveforms, 
we employ the numerical-relativity (NR) remnant fitting formulas (see, for example, Refs.~\cite{Healy:2014yta,Healy:2016lce,Jimenez-Forteza:2016oae,
Healy:2018swt,Hofmann:2016yih,Varma:2018aht} and references therein), 
from which the final \add{mass} $M_f$ and spin ${\bf S}_f$ of the remnant BH are consistently inferred 
for a given initial BH masses $m_{1,2}$ and dimensionless spin parameters 
$\vec{\chi}_{1,2}$ (in the inspiral phase) as
\footnote{\add{The mass and spin of the remnant BH in the remnant formulas are derived in the isolated horizon framework (see, for example, Ref.~\cite{Ashtekar:2004cn}), not obtained from the ringdown GWs. The latter is used only 
for checking the internal consistency of the formulas.}} 
\begin{equation}
\label{eq:M_f}
M_f = M_f(m_1,\,m_2,\,\vec{\chi}_1,\,\vec{\chi}_2) \,,
\quad 
\chi_f \equiv 
\frac{|{\bf S}_f|}{M_f^2} = \chi_f(m_1,\,m_2,\,\vec{\chi}_1,\,\vec{\chi}_2) \,.
\end{equation}
We then generate the accurate numerical data of $f_{\rm QNM}$ of the inferred remnant BH 
with the Black Hole Perturbation Club (B.H.P.C.) 
code~\cite{BHPC} to obtain $f_c$ and $Q$
\footnote{For accurate numerical data of $f_{\rm QNM}$, 
see also Ref.~\cite{Berti:2009kk}, Emanuele Berti's ``Ringdown'' website~\cite{Berti_QNM},
Ref.~\cite{Cook:2014cta} 
and the Black Hole Perturbation Toolkit~\cite{BHPT}.}.
In practice, it is also convenient to present $f_c$ and $Q$ 
by means of a compact analytical formula. 
Such a formula for the ($\ell=2,\,m=2,\,n=0$) mode is obtained in Ref.~\cite{Berti:2005ys} 
(by performing fits to the numerical QNM frequency data), and it reads 
\begin{equation}
\label{eq:QNM_fitting}
f_{c} = \frac{1}{2\pi M_f} \left[f_1+f_2\,(1-\chi_f)^{f_3}\right] \,,
\quad
Q = q_1+q_2\,(1-\chi_f)^{q_3} \,, 
\end{equation}
with 
$f_1 = 1.5251$, $f_2 = -1.1568$, $f_3 = 0.1292$,
$q_1 = 0.7000$, $q_2 = 1.4187$ and $q_3 = -0.4990$.

\subsection{Signal-to-noise ratio}
\label{subsec:SNR}

We estimate the SNR
of a complete IMR GW signal, 
making use of the simple frequency-domain, 
IMR ``amplitude'' model~\cite{Nakamura:2016hna}. 
This model is motivated by the waveform amplitude 
in the IMRPhenomB model~\cite{Ajith:2009bn} and it is given by 
\begin{equation}\label{eq:IMR}
{\rm IMR}(f) = {\cal A} \times \,
\begin{cases}
f^{-7/6} & {\rm for}~f < f_{\rm max} \,, \\ \\
f_{\rm max}^{-1/2} f^{-2/3} & {\rm for}~ f_{\rm max} \leq f < f_{\rm R}^{\rm r}/(1+z) \,, \\ \\
\displaystyle{
\frac{f_{\rm max}^{-1/2} \left[f_{\rm R}^{\rm r} / (1+z)\right]^{-2/3} \left[f_{\rm I}^{\rm r} / (1+z)\right]^2}
{\left\{f-\left[f_{\rm R}^{\rm r} / (1+z)\right]\right\}^2 + \left[f_{\rm I}^{\rm r} / (1+z)\right]^2}} 
& {\rm for}~ f_{\rm R}^{\rm r}/(1+z) \leq f \,,
\end{cases}
\end{equation}
where the overall constant ${\cal A}$ is chosen to be the (averaged) GW signal's amplitude\rem{s}  
in the inspiral phase~\eqref{def-A}. 
We set $f_{\rm max} = 1/[6^{3/2}\pi (1+z) m_t^{\rm r}]$
as the GW frequency at the innermost stable circular orbit (ISCO) 
of a test particle in the Schwarzschild spacetime with the total mass of BBH $m_t^{\rm r}$,
and $f_{\rm R}^{\rm r}$ and $f_{\rm I}^{\rm r}$ are the real and imaginary parts of the 
QNM frequency (i.e., $f_{\rm QNM}^{\rm r} = f_{\rm R}^{\rm r} + i\, f_{\rm I}^{\rm r}$)
of the fundamental $(n = 0)$, $\ell = 2 = m$ mode, 
which are determined by the final mass and spin of the remnant BH; 
the redshift dependence appears because the model parameters 
$(m_t^{\rm r},\,f_{\rm R}^{\rm r},\,f_{\rm I}^{\rm r})$ are all given 
in the source's rest frame 
(while the GW frequencies $f$ and $f_{\rm max}$ are in the observer frame). 
The averaged SNR (in the above sense) can be then obtained 
by 
\footnote{
For the given Fourier transform of a GW signal ${\tilde {h}}(f)$, 
the SNR can be written in terms of either ${\tilde {h}}(f)$ 
itself with the dimension of $1/{\rm Hz}$, 
the spectral density of the source amplitude \rem{is} 
$\sqrt{S_h(f)} \equiv 2 \sqrt{f}\, |{\tilde {h}}(f)|$
with the dimension of $1/\sqrt{\rm Hz}$, 
or 
the dimensionless characteristic strain 
$h_c(f) \equiv 2 f\, |{\tilde {h}}(f)|$~\cite{Moore:2014lga}\rep{.}{:} 
\begin{equation*}
\rho_{\rm ave}
=
\left(\int_{f_{\mathrm {in}}}^{f_{\mathrm {end}}} 
\frac{|2 {\tilde {h}} (f)|^2}{S_n(f)} df \right)^{1/2}
=
\left(\int_{f_{\mathrm {in}}}^{f_{\mathrm {end}}} 
\frac{|\sqrt{S_h(f)}|^2}{S_n(f)} \frac{df}{f} \right)^{1/2}
= 
\left(\int_{f_{\mathrm {in}}}^{f_{\mathrm {end}}} 
\frac{[h_c(f)]^2}{f\,S_n(f)} \frac{df}{f} \right)^{1/2} \,.
\end{equation*}
}
\begin{equation}
\label{rho-ave}
\rho_{\rm ave}
=
2 \, 
\left\{
\int_{f_{\mathrm {in}}}^{f_{\mathrm {end}}} 
\frac{[{\rm IMR}(f)]^2}{S_n(f)} df 
\right\}^{1/2} \,,
\end{equation}
where $S_n(f)$ is the noise PSD, 
and the frequency range $[f_{\mathrm {in}},\,f_{\mathrm {end}}]$ is 
determined by the GW detector with which we observe the GW signals. 

Note that the amplitude spectral density 
(i.e., the square root of the PSD of the source amplitude\rem{s})
is $\sqrt{S_h(f)} \equiv 2\sqrt{f}\, |{\rm IMR}(f)|$ in this model,  
and its track for the GW190521-like BBH is plotted in Figure~\ref{fig:N_curve} 
with the (square root of) the noise PSD $\sqrt{S_n(f)}$, 
assuming the median values of Table~\ref{table:GW190521} 
and the remnant formulas of Ref.~\cite{Healy:2014yta}.

\subsection{Multiband Fisher analysis}
\label{subsec:Fisher}

We approximate the variance (\add{i.e.,} uncertainty squared) associated
with the measurement of a set of signal parameters, 
making use of the standard Fisher matrix formalism. 
The Fisher information matrix for a single-band GW detector is defined by 
\begin{equation}
\Gamma_{a b}
\equiv 
\left. \left(\frac{\partial {\tilde h}}{\partial \theta_a} 
\left|
\frac{\partial {\tilde h}}{\partial \theta_b} \right.\right)
\right|_{ {\boldsymbol{\theta}} = {\boldsymbol{\theta}}_0}\,,
\end{equation}
where ${\tilde {h}}(f,\, {\boldsymbol {\theta}})$ is the frequency-domain GW signal 
described by the set of parameters ${\boldsymbol {\theta}}$, 
\add{and} ${\boldsymbol{\theta}}_0$ \rep{is}{are} the best-fit values of the binary parameters.
The bracket defines the noise-weighted inner product 
over the frequency range of $[f_{\mathrm {in}},\,f_{\mathrm {end}}]$~\cite{Finn:1992wt}  
\begin{equation}\label{eq:inner}
\left( a \mid b \right) 
\equiv 
2 \int_{f_{\mathrm {in}}}^{f_{\mathrm {end}}} 
\frac{{\tilde a}^* (f) {\tilde b}(f) + {\tilde b}^* (f) {\tilde a}(f)}
{S_n(f)} \, df \,,
\end{equation}
with asterisk `$*$', denoting the complex conjugation.
The inverse Fisher matrix defines the corresponding variance-covariance matrix 
$\Sigma^{a b} \equiv ( \Gamma_{a b} )^{-1}$.
In the limit of \rep{suitable}{suitably} high SNR~\cite{Vallisneri:2007ev}, 
the variance of the parameter $\theta^a$ \rep{are then}{is} given by 
\begin{equation}\label{eq:sta-error1}
\sigma_a^2 
= {\Sigma^{aa}}\,.
\end{equation}

In the case of the multiband analysis 
(to combine the information from, for example, aLIGO $+$ B-DECIGO), 
we simply construct a \rep{muitiband}{multiband} SNR and Fisher matrix 
by adding individual (averaged) SNRs and Fisher information matrices for each GW detector: 
\begin{equation}
\label{Fisher-M}
\rho_{\mathrm {tot}}^2
\equiv 
\sum_I (\rho_{\mathrm {ave}}^{I})^2\,,
\quad
{\Gamma}_{ab}^{\mathrm {tot}}
\equiv 
\sum_I {\Gamma}_{ab}^{I}\,, 
\end{equation}
where $\rho_{\mathrm {ave}}^{I}$ and $ {\Gamma}_{ab}^{I}$
are the averaged SNR and the Fisher matrix
for the $I$-th detector. 
The multiband variance-covariance matrix is defined by 
\begin{equation}
\label{variance-M}
 \Sigma^{a b}_{\mathrm {tot}} \equiv ( \Gamma_{a b}^{\mathrm {tot}} )^{-1}\,, 
\end{equation}
and the variance of $\theta^a$ is then obtained 
by $\sigma^2_a = \Sigma^{a a}_{\mathrm {tot}}$.

\section{Parameter estimation errors via multiband observation}
\label{sec:params-err}

In this section, we summarize the parameter estimation errors
\add{of the inspiral and ringdown phases} for a \rep{non-precessing}{nonprecessing}, spinning, GW190521-like BBH system, 
using the multiband GW network (LISA, B-DECIGO, aLIGO and ET) 
detailed in Section~\ref{subsec:detectors}. 
We follow Refs.~\cite{Nakano:2015uja,Isoyama:2018rjb} 
in our treatment of the Fisher matrix calculation for BBH GW signals
\footnote{The setup here is slightly different from Ref.~\cite{Isoyama:2018rjb}; 
i) we will assume the $5\,$yr observation, rather than $4\,$ yr observation, 
and ii) we will use the new LISA sensitivity curve proposed 
by Ref.~\cite{Cornish:2018dyw} and displayed in Eq.~\eqref{S-LISA}, 
not the earlier eLISA sensitivity curve presented 
in Ref.~\cite{Babak:2017tow}.},
and we continue to neglect the contribution from the merger GW signal
\footnote{\add{Using our IMR amplitude model in Eq.~\eqref{eq:IMR}, 
we have the (averaged) merger SNR of $17.0$, $1.97 \times 10^2$ and $3.42$
for aLIGO, ET and B-DECIGO, respectively. The addition of these contributions 
to the signal analysis will improve the parameter estimations.}}; 
\add{for the reason that one cannot separate the inspiral, merger and ringdown phases cleanly in the strict sense 
although we have presented the simple IMR amplitude model in Eq.~\eqref{eq:IMR}.
When the merger contribution is introduced into the inspiral or ringdown signal analysis, it causes some bias in the inspiral or ringdown parameter estimation. 
For example, the merger--ringdown waveform is parametrized by the binary parameters, 
not solely by the remnant BH parameters after the merger. Similarly, in the inspiral--merger waveform, the ``late'' merger phase can be described by the overtones of QNMs~\cite{Giesler:2019uxc}, i.e., the remnant BH parameters.
Thus, we shall perform the inspiral--ringdown consistency test of GR without the merger contribution in Section~\ref{subsec:testingGR}.}

\subsection{Setup of Fisher analysis}
\label{subsec:num-setup}

We set the default frequency interval of each GW detector 
$[f_{\mathrm {low}},\, f_{\mathrm {up}}]$ as 
$[10.0,\, 3.0 \times 10^3]~ {\rm Hz}$ (aLIGO), 
$[2.0 ,\, 3.0 \times 10^3]~ {\rm Hz}$ (ET), 
$[0.01,\, 1.0 \times 10^2]~ {\rm Hz}$ (B-DECIGO), 
and 
$[1.0 \times 10^{-4},\, 1.0]~ {\rm Hz}$ (LISA), respectively.  
We shall adopt the $T_{\rm obs} = 5$-year observation time, 
and assume that the binary merges at the GW frequency of the Schwarzschild ISCO,
$f_{\rm ISCO} = 1/[6^{3/2}\pi (1+z) m_t^{\rm r}]$. 
In this setup, the minimum frequency of the GW signal is 
\begin{equation}
f_{\rm min} =  9.24 \times 10^{-3}\,(1+z)^{-5/8} 
\left(\frac{{\cal M}^{\rm r}}{65.1M_\odot}\right)^{5/8}
\left(\frac{5\, {\rm yr}}{T_{\rm obs}}\right)^{3/8}\, {\rm Hz}\,,
\end{equation}
where ${\cal M}^{\rm r}$ is the chirp mass in the source's rest frame, 
normalized to that of GW190521. 
Therefore, the GW signal\rem{s} observed by each GW detector is truncated 
at the corresponding initial frequency 
$f_{\rm in} \equiv \max(f_{\rm min},\,f_{\rm low})$ 
as well as the end frequency 
$f_{\rm end} \equiv \min(f_{\rm ISCO},\,f_{\rm up})$; 
recall Figure~\ref{fig:N_curve}.

The parameters of the \rep{inspial}{inspiral} waveform~\eqref{eq:hf} are  
\begin{equation}
\label{eq:insp}
\boldsymbol{\theta}_{\mathrm {Insp}} 
= (\rem{f_0\,}t_c,\, \Psi_c,\, \ln m_t,\, \nu,\, \chi_s,\, \chi_a)\,,
\end{equation}
where \rem{$f_0 = 1.65 \,{\mathrm {Hz}}$ at which B-DECIGO is most sensitive,
and} we define the symmetric and anti-symmetric combinations of BH spins by  
$\chi_s \equiv {(\chi_1 + \chi_2)}/{2}$ 
and
$\chi_a \equiv {(\chi_1 - \chi_2)}/{2}$
with the component (aligned) BH spins 
$\chi_{1,2} \equiv |\vec{\chi}_{1,2}|
\equiv {|{{\bf S}_{1,2}|}}/{m_{1,2}^2}$.
At the same time, the parameters of the ringdown waveform~\eqref{eq:f-RDwave} are  
\begin{equation}
\label{eq:ring}
\boldsymbol{\theta}_{\mathrm {Ring}} 
= (t_0,\, \phi_0,\, f_c,\, Q)\,.
\end{equation}
It should be noted that the amplitude parameters are left out from the set of 
our independent parameters both in Eqs.~\eqref{eq:insp} and~\eqref{eq:ring}. 
They are entirely uncorrelated with other parameters $\theta_a$ 
because the variance-covariance matrix $\Sigma^{a\,b}$ gives 
the variance $\sigma_{\ln {\cal A}}^2 = \rho_{\rm ave}^{-2}$ 
and the correlation $c^{{\ln {\cal A}},\,a} 
\equiv 
\Sigma^{{\ln {\cal A}},\,a} / 
(\sigma_{\ln {\cal A}}\,\sigma_{a} ) = 0$ 
for the inspiral GW signal~\eqref{eq:hf} 
(see, for example, Ref.~\cite{Poisson:1995ef}), 
and similar for the ringdown GW signal (if we explicitly introduce the amplitude 
to the normalized waveform of Eq.~\eqref{eq:f-RDwave}).
For simplicity, we consider that all (other) parameters are unconstrained.

The best fit \rem{parameters} values of \rem{the} inspiral \add{parameters} are given 
by the median values in Table~\ref{table:GW190521} with $t_c = 0.0 = \Psi_c$, 
while those of the ringdown \rep{GW signal}{parameters} are assumed to be $\phi_0 = 0.0 = t_0$
\footnote{The parameter estimation errors are independent of the value of the initial time $t_0$. 
In our frequency-domain, single-mode ringdown waveform in Eq.~\eqref{eq:f-RDwave}, 
the $t_0$ dependence is factorized as ${{\rm e}^{2\,i\pi\,f\,t_0}}$ 
and it does not contribute to the noise-weighted inner product in Eq.~\eqref{eq:inner}.
On the other hand, the estimation errors depend on the initial phase $\phi_0$ weakly~\cite{Nakano:2004ib}.
}, 
and
\begin{equation}
\label{eq:median-fQ}
(f_c,\, Q) = (85.061\,{\rm Hz},\,4.8354)\,,
\end{equation}
for the (observer-frame) central frequency and quality factor. 
The values in Eq.~\eqref{eq:median-fQ} are obtained via Eq.~\eqref{eq:f_Q} 
for a given QNM frequency data of the remnant BH, 
assuming that the remnant mass $M_f$ and spin $\chi_f$ of the final BH are inferred 
via the remnant formula in Eq.~\eqref{eq:M_f} with the parameters of each component BH 
in Table~\ref{table:GW190521}. 
Specifically, we use the remnant formula provided by Ref.~\cite{Jimenez-Forteza:2016oae}, 
and we quote
\begin{equation}
\label{eq:remnants}
\left(\frac{M_f}{m_t},\, \chi_f\right) = (0.90356,\,0.88269)\,.
\end{equation}
\add{
Note that these values are different from the mass and spin 
$({M_f},\, \chi_f) = (142^{+28}_{-16}\,M_{\odot},\,0.72^{+0.09}_{-0.12})$ 
of the remnant BH reported in the LIGO/Virgo GW190521 detection paper~\cite{Abbott:2020tfl};  
we have assumed that the individual spins of GW190521-like BBH 
are nonprecessing, and completely aligned to the orbital angular momentum for simplicity 
while the observed GW190521 is actually considered to be a precessing BBH.
} 
The associated QNM frequency data of this remnant BH is then generated 
by the B.H.P.C. code~\cite{BHPC}, yielding the results in Eq.~\eqref{eq:median-fQ}.

\rep{To see the benefit of the multiband GW observation, finally, 
we introduce the normalized root-mean-square errors as}
{
The root-mean-square of parameter estimation errors scales like the inverse of SNR, $\sim 1 / \rho$, and it depends on both $\rho$ and bandwidth over which the SNR is accumulated. To see the benefit of having wider bandwidth due to the multiband observation, finally, we introduce the normalized root-mean-square error as
}
\begin{equation}\label{hat-theta}
\delta {\boldsymbol {\hat \theta}}
\equiv 
\rho\,{\boldsymbol {\sigma}} \,,
\end{equation}
and shall display our error estimations in terms of $\delta {\boldsymbol {\hat \theta}}$ 
with the total averaged SNR $\rho_{\rm ave}^{\rm tot}$ accumulated over \rep{multibands}{multi-frequency bands}.

\subsection{Result: Inspiral phase}
\label{subsec:insp}

In Table~\ref{table:sta_err_GW190521},
we present the parameter estimation errors of mass parameters $(m,\,\nu)$ 
and spin parameters $(\chi_s,\,\chi_a)$ 
for the GW190521-like BBH inspiral
in various combinations of ground/space-based GW observatories 
(but suppressing those of $t_c$ and $\Psi_c$). 
Here, it should be noted that only the inspiral phase
is analyzed here, and the merger--ringdown phase
which contributes to the SNR for ground-based observatories, is ignored \add{; recall the footnote 9}.
Therefore, the SNR for aLIGO quoted here is much smaller 
than the observed LIGO/Virgo network SNR of $14.5$~\citep{Abbott:2020tfl}.

In the single-band case, as seen in Figure~\ref{fig:N_curve},
the inspiral GW signal of \add{the} GW190521-like BBH\rem{s} \rep{are}{is} best observed by B-DECIGO 
because it can cover both the early (1\,yr before merger) 
and late (around the ISCO) phases. 
However, B-DECIGO observation alone \rep{does}{is} not enough benefit to discern the spin parameters. 
\rem{
Also, interestingly, the normalized errors of LISA observation is better than those of ET.
Although the SNR of LISA is one order of magnitude smaller than that of B-DECIGO and ET 
(and it would be actually too small compared to the `realistic' detection 
SNR threshold of $\sim 15$~\cite{Moore:2019pke}),  
the longer observation period ($\sim\,$ years) can reduce the normalized uncertainty.}

In the multiband cases, thanks to the wider bandwidth, 
the observation with B-DECIGO and ET gives a factor of $2$ improvement 
in all the parameter estimation, even if we observe only the inspiral GW signal. 
This is further refined if we combine the data from LISA, 
forming a triple-band network
\add{(although the estimated LISA SNR of $\sim 2.68$ is likely too small compared 
to the `realistic' detection SNR threshold of $\sim 15$~\cite{Moore:2019pke})}.
\add{Importantly, the normalized errors of BH spins 
($\delta {\hat \chi_s} / \chi_s$,\, $\delta {\hat \chi_a} / \chi_a$)
=
($2.52$,\,$5.71 \times 10^{2}$) in this best case imply that 
the magnitudes of individual BH spins $\sigma_{\chi}$ can be recovered 
with the fractional statistical errors 
(i.e., now accounting for the total multiband SNR of 
$\rho_{\rm ave}^{\rm tot} = 6.52 \times 10^{1}$ to the uncertainties; recall Eq.~\eqref{hat-theta})
($\sigma_{\chi_1} / \chi_1$,\, $\sigma _{\chi_2} / \chi_2$)
=
($2.43 \times 10^{-1}$,\,$2.51 \times 10^{-1}$).
}
\rem{Our result therefore implies that the {multibanding}}
\add{The joint} LISA, B-DECIGO and ET observatories would be only viable network 
to measure all the binary parameters of GW190521-like BBH system \add{in the inspiral phase}, 
including BH's component spins.

\begin{table}[ht]
\caption{Parameter estimation errors of mass parameters $(m,\,\nu)$ 
and spin parameters $(\chi_s,\,\chi_a)$ 
for \add{the} GW190521-like quasicircular BBH inspiral, 
normalized to the total multiband SNR $\rho_{\rm tot}$; 
for example, the result of ``BD'' in ``BD $+$ aLIGO'' is normalized 
to $\rho_{\rm tot} = 5.93 \times 10^1$. 
We assume that the true binary parameters are given 
by the median values of Table~\ref{table:GW190521},
i.e., $m_t=151.0\,M_{\odot}$, $\nu=0.246$,
$\chi_s=0.71$ and $\chi_a=0.02$.
Since the source location can be determined
well by B-DECIGO (BD here)~\cite{Nakamura:2016hna, Nair:2018bxj}, 
we fix $1+z = 1.82$.
Note that only the inspiral SNR of aLIGO is $1.76$,
which is too small to give meaningful estimation errors. 
Also any \add{normalized} estimation errors 
$\delta {\boldsymbol {\hat \theta}} > 1.0 \times 10^6$ 
are discarded from this table. 
}
\centering
\begingroup
\renewcommand{\arraystretch}{1.2} 
\scalebox{0.8}[0.8]{
\begin{tabular}{lc|cccc}
\toprule
GW detector  \quad & \quad SNR & 
\quad ${\delta {\hat m}_t} / m_t$ & \quad ${\delta {\hat \nu}} / \nu$ & 
\quad $\delta {\hat \chi_s} / \chi_s$  & 
\quad $\delta {\hat \chi_a} / \chi_a$ \\
\midrule
\multicolumn{6}{l}{BD + aLIGO}\\
\midrule
BD \quad & \quad $5.92 \times 10^{1}$ &
\quad $3.51 \times 10^{-1}$ & \quad $5.84 \times 10^{-1}$ & 
\quad $4.91$ & \quad $1.12 \times 10^{3}$ \\ 
aLIGO \quad & \quad $1.76$ &
\quad $\cdots$ & \quad $\cdots$ & 
\quad $\cdots$ & \quad $\cdots$ \\ 
BD + aLIGO  \quad & \quad $5.93 \times 10^{1}$ &
\quad $3.44 \times 10^{-1}$ & \quad $5.73 \times 10^{-1}$ & 
\quad $4.73$ & \quad $1.08 \times 10^{3}$ \\ 
\midrule
\multicolumn{6}{l}{BD + ET}\\
\midrule
ET \quad & \quad $2.72 \times 10^{1}$ &
\quad $3.70 \times 10^{2}$ & \quad $6.00 \times 10^{2}$ & 
\quad $5.86 \times 10^{3} $ & \quad $\cdots$ \\ 
BD + ET  \quad & \quad $6.52 \times 10^{1}$ &
\quad $2.88 \times 10^{-1}$ & \quad $4.80 \times 10^{-1}$ & 
\quad $3.07$ & \quad $7.01 \times 10^{2}$ \\ 
\midrule
\multicolumn{6}{l}{LISA + BD + ET}\\
\midrule
LISA  \quad & \quad $2.68$ &
\quad $1.73 \times 10^{1}$ & \quad $2.89 \times 10^{1}$ & 
\quad $2.37 \times 10^{3} $ & \quad $\cdots$ \\ 
LISA + BD \quad & \quad $5.93 \times 10^{1}$ &
\quad $1.83 \times 10^{-1}$ & \quad $3.04 \times 10^{-1}$ & 
\quad $4.01$ & \quad $9.11 \times 10^{2}$ \\ 
LISA + BD + ET \quad & \quad $6.52 \times 10^{1}$ &
\quad $1.45 \times 10^{-1}$ & \quad $2.41 \times 10^{-1}$ & 
\quad $2.52$ & \quad $5.71 \times 10^{2}$ \\ 
\bottomrule
\end{tabular}
}
\endgroup
\label{table:sta_err_GW190521}
\end{table}

\subsection{Result: Ringdown phase}
\label{subsec:RD}

In Table~\ref{table:sta_err_RD},
we show the parameter estimation errors of the central frequency $f_c$ 
and quality factor $Q$ (but suppressing those of $t_0$ and $\phi_0$) 
for the ringdown phase of \add{the} GW190521-like BBH in ground-based observatories.
For simplicity, we estimate the \rep{ringdonw}{ringdown} amplitude and associated SNR 
via the IMR amplitude model in Eq.~\eqref{eq:IMR}
\footnote{We should note that this approximation is likely too raw 
because the ringdown amplitude strongly depends on the starting time $t_0$ 
of the ringdown phase, which is difficult to determine in practice.}.
We see that ET will be able to measure \add{the} QNM frequency \add{of the remnant BH} 
with the statistical error $\sim 10^{-3}$. 
At the same time, however, the difference in the normalized errors 
$\delta {\hat f}_c$ and $\delta {\hat Q}$ between aLIGO and ET observations \rep{are}{is} not so evident.  
We speculate \add{that} it arises from the difference in the spectrum shapes (not the overall amplitudes)   
because the ringdown waveform~\eqref{eq:f-RDwave} is narrow banded in the frequency domain.

\begin{table}[ht]
\caption{Parameter estimation errors of the central frequency $f_c$ 
and the quality factor $Q$ of \add{the} GW190521-like \rep{BBH}{remnant BH}, 
only using the ringdown GW signal. 
The errors are normalized to the SNR of each corresponding \rep{observatories}{detector}.}
\centering
\begingroup
\renewcommand{\arraystretch}{1.2} 
\scalebox{0.8}[0.8]{
\begin{tabular}{lc|cc}
\toprule
GW detector  \quad & \quad SNR & 
\quad ${\delta {\hat f}_c} / f_c$ & \quad ${\delta {\hat Q}} / Q$ \\
\midrule
aLIGO \quad & \quad $1.08 \times 10^{1}$ &
\quad $3.42 \times 10^{-1} $ & \quad $2.71 \times 10^{-1}$ \\ 
ET  \quad & \quad $1.47 \times 10^{2}$ &
\quad $3.66 \times 10^{-1}$ & \quad $2.86 \times 10^{-1}$ \\ 
\bottomrule
\end{tabular}
}
\endgroup
\label{table:sta_err_RD}
\end{table}

\section{The implications for tests of GR via multiband observation}
\label{sec:implication}

In this section, we explore what extent the multiband observation
of \add{the} GW190521-like BBH system discussed in Section~\ref{sec:params-err} 
could improve tests of GR. 
A handful of tests have been already formulated and performed with merging BBH systems 
(see, for example, 
Refs.~\cite{TheLIGOScientific:2016src,LIGOScientific:2019fpa,Abbott:2020jks,Carson:2019yxq}), 
and we follow the (very) simple tests proposed  
by Nakano et al.~\cite{Nakano:2015uja,Nakano:2016sgf}.

\subsection{A consistency test of GR 
with the inspiral and ringdown GW signals}
\label{subsec:testingGR}

One possible test of GR with a BBH system is to establish the consistency 
of the mass and spin of the final remnant BH determined 
by two different parts of the GW signals. 
Thanks to the recent advancement in NR simulations 
of BBH systems~\cite{Pretorius:2005gq, Campanelli:2005dd, Baker:2005vv} 
(see also Refs.~\cite{Jani:2016wkt,Healy:2017psd,Healy:2019jyf,Boyle:2019kee,Healy:2020vre}), 
one can infer these values from the initial component masses 
and spins measured from the inspiral GW signal (in the low-frequency band), 
making use of the NR fitting formulas for the remnant properties of the final BH; 
recall Section~\ref{subsec:GW}.
At the same time, they are directly estimated from the succeeding merger--ringdown GW signal 
(in the high-frequency band). 
This type of test is now known as the ``IMR consistency test''~\cite{Ghosh:2016qgn,Ghosh:2017gfp}. 
By formulation, multiband observations of heavy BBH mergers such as GW190521-like BBH systems
will be ``golden binaries''~\cite{Hughes:2004vw} of such an IMR consistency test. 

Given that the early inspiral and late ringdown GW signals 
will be best observed in a different frequency band 
(such as ``B-DECIGO $+$ aLIGO'' network etc.), 
we here perform the multiband version of the ``inspiral--\rep{ringdwon}{ringdown} \rep{(IR)}{consistency} test'' formulated 
by Nakano et al.~\cite{Nakano:2015uja} 
(see also Refs.~\cite{Hughes:2004vw,Luna:2006gw}), 
solely using the inspiral and \rep{ringdwon}{ringdown} parts, 
and test the consistency of GR across the merger part, 
which is \add{a} highly dynamical phase in a strong-field regime.
We estimate the statistical errors on $M_f$ and $\chi_f$ from the inspiral GW signal 
by using the (\rep{normalised}{normalized}) statistical errors 
${\delta {\boldsymbol {\hat \theta}}}$ in Table~\ref{table:sta_err_GW190521} 
and applying a standard variance propagation of non-linear functions to 
the specific NR remnant formulas 
(``UIB formulas'')~\cite{Jimenez-Forteza:2016oae} 
publicly available in LALInference~\cite{Veitch:2014wba, lalsuite} 
\footnote{
We ignore the systematic bias due to our specific choice 
of the NR remnant formulas, for simplicity. 
See, for example, Refs.~\cite{Ghosh:2017gfp,Boyle:2019kee} for details. 
}.
%
The errors from the ringdown GW signal are estimated from Table~\ref{table:sta_err_RD}, 
through the dependence of $f_c$ and $Q$ on $(M_f,\,\chi_f)$
\footnote{Technically, this procedure requires the evaluation 
of the partial derivatives $(\partial f_c / \partial \chi_f)_{M_f}$ etc. 
to compute the variance propagation. 
We construct the numerical function of QNM frequencies $ M_f\,f_{\rm QNM}(\chi_f)$ 
with the B.H.P.C. code~\cite{BHPC} around the best fit values of Eq.~\eqref{eq:remnants}, 
from which these derivatives are extracted.}.

In Table~\ref{table:IMR_test}, we present the parameter estimation
errors of the remnant mass $M_f$ and spin $\chi_f$ from both the inspiral and ringdown phases.
For the reference, we also present the same errors of \add{the} GW150914-like BBH 
with the redshifted component masses $(m_1 ,\,m_2) = (30\,M_\odot,\,40\,M_\odot)$, 
\rem{the component} spin magnitudes \rep{$(\chi_1 ,\,chi_2) = (0.9, 0.7)$}{$(\chi_1 ,\,\chi_2) = (0.9, \,0.7)$},  
and the luminosity distance $D_L=0.4\,{\rm Gpc}$ (i.e., the redshift $z \sim 0.085$)
\footnote{This system was analyzed as ``System B'' 
in Ref.~\cite{Isoyama:2018rjb}.}.
We see that \rep{multiband}{the} ``LISA $+$ B-DECIGO $+$ ET'' \add{observation} allows us the IMR consistency test 
at the sub-percent precision both for \add{the} GW150914-like, and GW190521-like BBH systems.
The result of \add{the} GW190521-like BBH system is slightly worse than the case of 
\add{the} GW150914-like BBH system because the luminosity distance of GW190521 ($D_L=5.3\,{\rm Gpc}$) 
is much larger than that of GW150914
($D_L=0.4\,{\rm Gpc}$)~\cite{Abbott:2016blz,TheLIGOScientific:2016wfe,Abbott:2016izl}.
Also, the GW190521-like BBH system has
the central frequency $f_c \sim 85\,{\rm Hz}$ lower 
than that of the GW150914-like BBH system ($f_c \sim 350\,{\rm Hz}$), 
missing the most sensitive frequency of aLIGO and ET $\sim 250\,{\rm Hz}$.
If a GW190521-like BBH was observed at the same distance as GW150914, 
the accuracy of its \rep{IMR}{inspiral--ringdown} consistency test would be at the level of $\sim O(0.01\%)$.

\begin{table}[ht]
\caption{Parameter estimation errors of the remnant mass and spin of GW190521-like 
and GW150914-like BBH systems, using only the ringdown GW signal as well as the inspiral GW signal 
inferred via the NR remnant fitting formulas; 
the best fit values $(M_f / m,\, \chi_f) \sim (0.904,\,0.883)$
for the GW190521-like BBH (recall Eq.~\eqref{eq:remnants}) and
$(M_f / m,\, \chi_f) \sim (0.891,\,0.898)$ for the GW150914-like BBH.
Here, BD is an abbreviation for B-DECIGO. 
The results in the single band are normalized to the SNR 
for a given GW observatory. In the multiband case they are normalized 
to the total SNR of $\rho_{\rm {LISA + BD + ET}}$.}
\centering
\begingroup
\renewcommand{\arraystretch}{1.2} 
\scalebox{0.8}[0.8]{
\begin{tabular}{l|ccc|ccc}
\toprule
 & \multicolumn{3}{c|}{GW190521-like BBH}
 & \multicolumn{3}{c}{GW150914-like BBH} \\
GW detector  \quad & \quad SNR & 
\quad ${\delta M_f} / M_f$ & \quad ${\delta \chi_f} / \chi_f$ &
\quad SNR &
\quad ${\delta M_f} / M_f$ & \quad ${\delta \chi_f} / \chi_f$ \\
\midrule
\multicolumn{7}{l}{Single band: Ringdown GW signal}\\
\midrule
aLIGO \quad & \quad $1.08 \times 10^{1}$ &
\quad $3.09 \times 10^{-1} $ & \quad $7.47 \times 10^{-1}$
& \quad $1.78 \times 10^{1}$ &
\quad $3.05 \times 10^{-1} $ & \quad $6.47 \times 10^{-1}$ \\ 
ET  \quad & \quad $1.47 \times 10^{2}$ &
\quad $3.31 \times 10^{-1}$ & \quad $7.90 \times 10^{-1}$
& \quad $2.66 \times 10^{2}$ &
\quad $3.26 \times 10^{-1}$ & \quad $6.60 \times 10^{-1}$ \\ 
\midrule
\multicolumn{7}{l}{Single band: inspiral GW signal}\\
\midrule
BD \quad & \quad $5.92 \times 10^{1}$ &
\quad $3.31 \times 10^{-1}$ & \quad $9.63 \times 10^{-1}$
& \quad $2.51 \times 10^{2}$ &
\quad $3.19 \times 10^{-1}$ & \quad $9.43 \times 10^{-1}$ \\ 
\midrule
\multicolumn{7}{l}{Multiband: inspiral GW signal}\\
\midrule
BD + ET  \quad & \quad $6.52 \times 10^{1}$ &
\quad $3.54 \times 10^{-1}$ & \quad $5.90 \times 10^{-1}$
& \quad $5.18 \times 10^{2}$ &
\quad $1.07 \times 10^{-1}$ & \quad $4.48 \times 10^{-1}$ \\ 
LISA + BD  \quad & \quad $5.93 \times 10^{1}$ &
\quad $2.24 \times 10^{-1}$ & \quad $7.32 \times 10^{-1}$
& \quad $2.51 \times 10^{2}$ &
\quad $3.10 \times 10^{-1}$ & \quad $8.96 \times 10^{-1}$ \\ 
LISA + BD + ET  \quad & \quad $6.52 \times 10^{1}$ &
\quad $1.75 \times 10^{-1}$ & \quad $4.98 \times 10^{-1}$
& \quad $5.18 \times 10^{2}$ &
\quad $1.06 \times 10^{-1}$ & \quad $4.33 \times 10^{-1}$ \\
\bottomrule
\end{tabular}
}
\endgroup
\label{table:IMR_test}
\end{table}

\subsection{A simple test of the remnant compact object with quasinormal modes}
\label{subsec:testing-rem}

Another simple test of GR is to bracket whether the remnant object should be 
a BH predicted by GR or not, making use of the parameter estimation errors 
of \rep{quasinormal mode (QNM)}{QNM} frequencies ($f_{\rm QNM} = f_{\rm R} + i\, f_{\rm I}$)
obtained from \rem{the} ringdown GW signals~\cite{Nakano:2015uja}.

\begin{figure}[ht]
\centering
\includegraphics[width=0.49\textwidth]{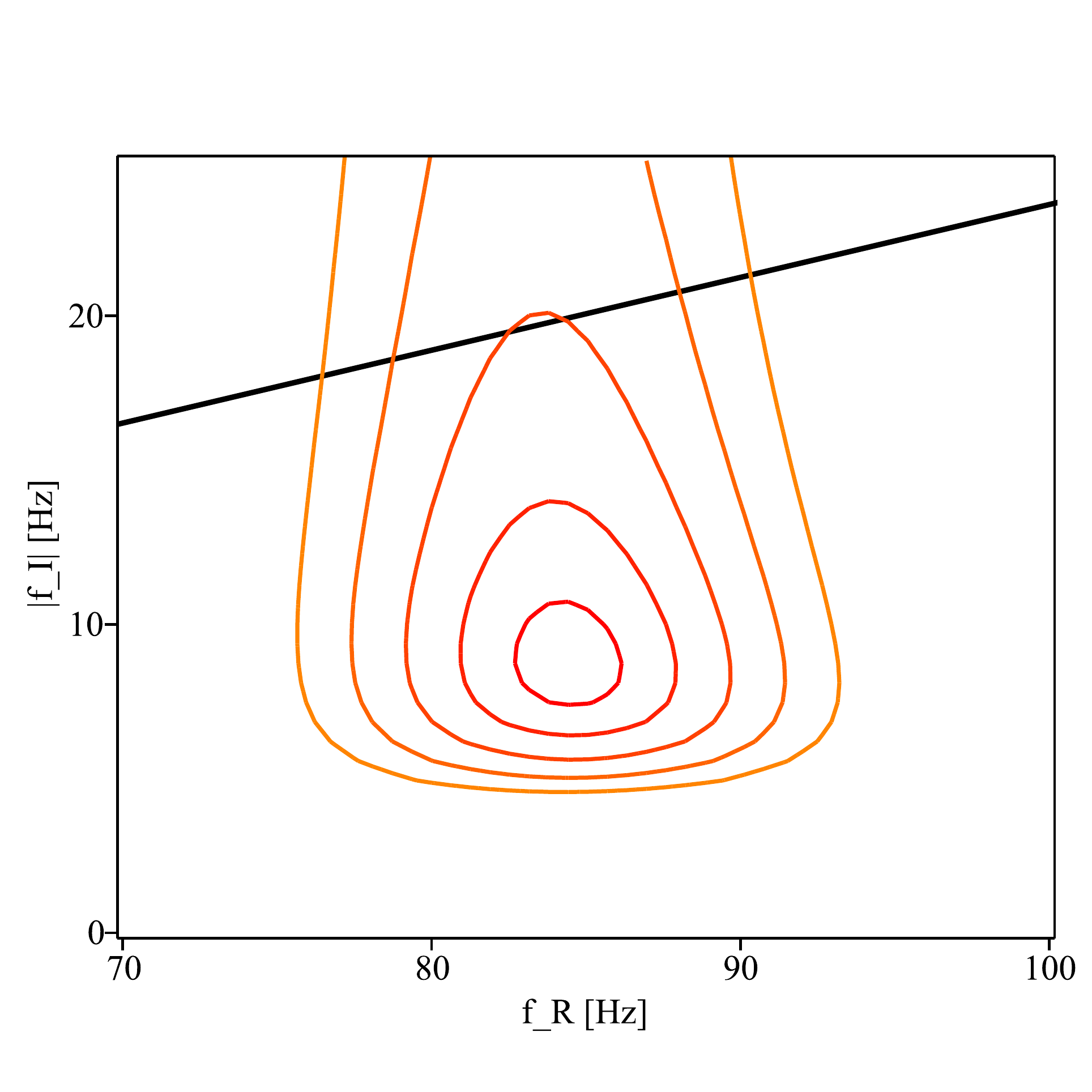}
\includegraphics[width=0.49\textwidth]{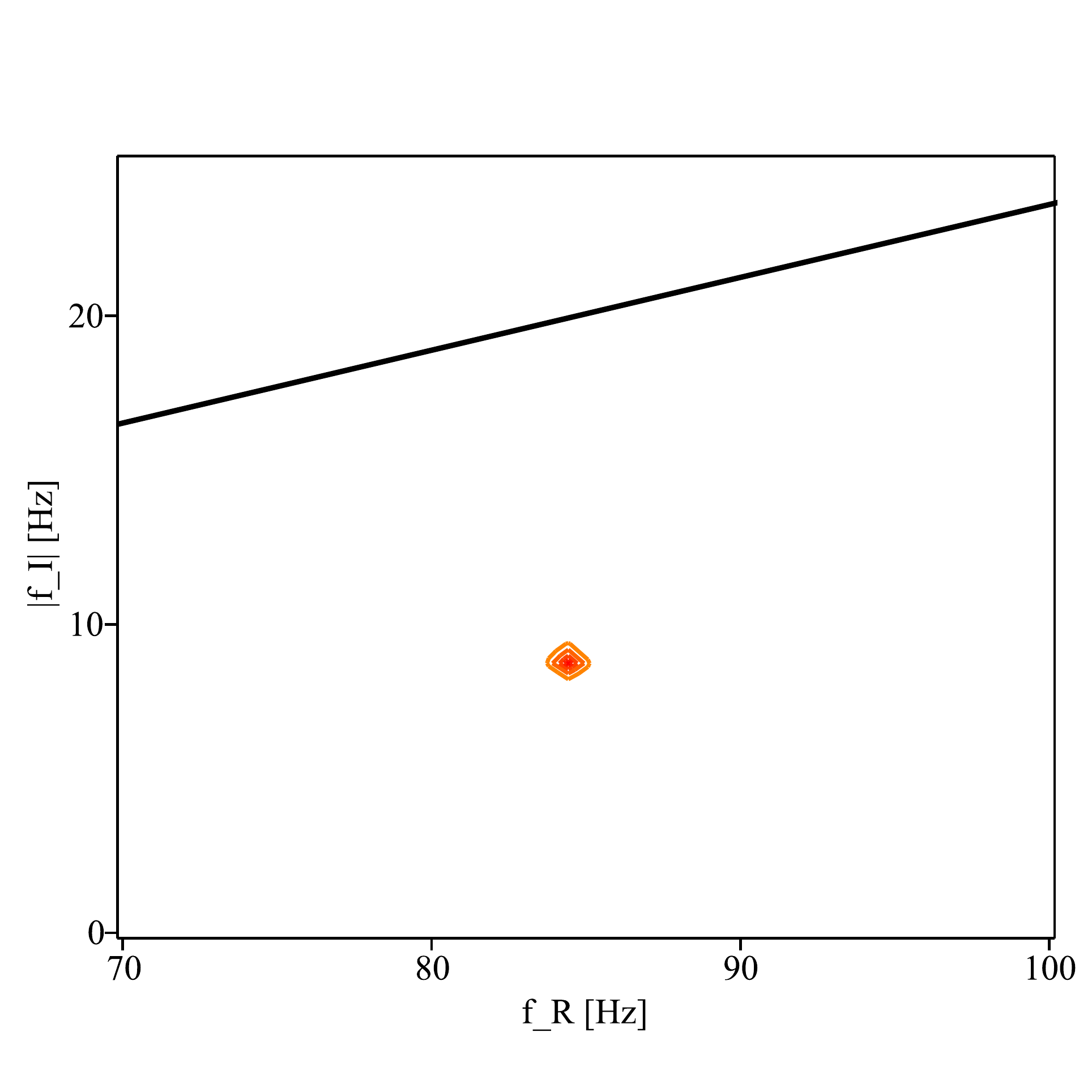}
\caption{Simple remnant test in the ($f_{\rm R},\,f_{\rm I}$) plane. 
The QNM frequency is given in the observer frame, 
and we assume the GW190521-like ringdown GW signals 
with $(M_f/m, \chi_f) \sim (0.904, 0.883)$~\eqref{eq:remnants}.
The figure shows the parameter estimation \rep{erro}{error} with 
the aLIGO noise curve (left, SNR$\,=1.08 \times 10^{1}$)
and the ET one (right, SNR$\,=1.47 \times 10^{2}$).
The black lines are the Schwarzschild limit of $|f_{\rm I}|/f_{\rm R} \approx 0.236$.
The colored ellipses show the contours of
$1\sigma,\,2\sigma,\,3\sigma,\,4\sigma,$ and $5\sigma$.
To obtain these two dimensional plots,
the time and phase parameters ($t_0,\,\phi_0$) have been marginalized out.}
\label{fig:simple_QNM}
\end{figure}

Figure~\ref{fig:simple_QNM} plots the $1\sigma,\,2\sigma,\,3\sigma,\,4\sigma,$ 
and $5\sigma$ error contours on the fundamental $(n = 0)$, $\ell = 2 = m$ mode 
of the QNM frequency in the ($f_{\rm R},\,f_{\rm I}$) plane,  
in the case of \add{the} GW190521-like BBH system observed by aLIGO (left) and ET (right).
The errors are estimated through the results in Table~\ref{table:sta_err_RD} 
about the parameter estimation error on the ringdown GW signal 
(after $t_0$ and $\phi_0$ being marginalized out), 
and the outermost contour in each panel shows the $5\sigma$ error.  
This black lines in each panel\rem{s} depict the Schwarzschild limit of 
${|f_{\rm I}|}/{f_{\rm R}}$, 
which may be obtained by setting $\chi_f=0$ in Eq.~\eqref{eq:QNM_fitting} 
with Eq.~\eqref{eq:f_Q} 
\footnote{
This is marginally consistent with the exact QNM frequency 
of the fundamental $(n = 0)$, $\ell = 2 = m$ mode in the Schwarzschild limit.   
For example, the B.H.P.C. code gives ${|f_{\rm I}|}/{f_{\rm R}} = 0.23808 \dots$, 
and the difference is negligibly small \rem{for our raw analysis} here.},
\begin{equation}
\label{eq:QNM-Sch}
\frac{|f_{\rm I}|}{f_{\rm R}} \approx 0.236 \,.
\end{equation}
The key point of this test is that the top-left side of the black line 
becomes the prohibited region in GR, i.e., for the $(n = 0)$, $\ell = 2 = m$ mode, 
the QNM frequencies ($f_{\rm R},\,f_{\rm I}$) of any rotating Kerr BHs 
must sit in the bottom-right side of the black line.

In the aLIGO case, due to the low SNR ($=10.8$), the parameter estimation errors already go 
beyond the Schwarzschild limit at the $3\sigma$ level, 
and we cannot confirm whether the remnant object is a BH predicted by GR
with the $5\sigma$ level. For such low SNR events, 
the ``coherent mode stacking method''~\cite{Yang:2017zxs} will be useful.
On the other hand, thanks to \rem{the} the high SNR ($=147$) in the ET case, 
this simple test can necessarily confirm that the remnant object is a GR-predicted BH
\footnote{The ($M_f,\,\chi_f$) plane has been discussed as Figure 5 
in Ref.~\cite{Abbott:2020tfl}.
The remnant BH spin is restricted to $0 \leq \chi_f < 1$ in the analysis. 
Therefore, the simple test presented in this paper is not applicable.}.

\section{Summary and Discussion}
\label{sec:discussion}

This work underlines the multiband observation of the GW190521-like 
nonprecessing, quasicircular ``intermediate-mass'' BBH 
with LISA and B-DECIGO in the low-frequency band, 
combined it with aLIGO and ET in the high-frequency band. 
Our first result of the parameter estimation errors was
displayed in Tables~\ref{table:sta_err_GW190521} and~\ref{table:sta_err_RD}; 
the statistical errors of the binary's mass parameters 
by B-DECIGO observation will be $\sim 10^{-2}$, 
and the \rep{multibanding}{multiband observation with} LISA, B-DECIGO and ET will further improve them 
to a factor of $2$, even allowing the statistical errors of component BH spins 
at the accuracy level of $\sim 10^{-1}$. 
Based on the ringdown analysis, ET will measure the QNM frequency about $O(0.1)\%$ precision. 
Our second result of GR tests was presented in Table~\ref{table:IMR_test} 
as well as \rep{Figures.~\ref{fig:simple_QNM} and~\ref{fig:NNT1}}{Figure~\ref{fig:simple_QNM}}. 
We showed that the multiband observation of \add{the} GW190521-like BBH system 
by LISA, B-DECIGO and ET would perform the inspiral--\rem{(merger)--}ringdown consistency test 
at the level of percent-precision. 
\rem{We also argued that the accurate QNM frequency measured by ET would be able to verify 
the remnant compact object to be a Kerr BH predicted by GR, proving its ergoregion.}

The main point of our analysis is that there is the principal advantage 
of measuring GW190521-like BBH systems using \textit{the full GW spectrum} 
from milli-Hz to deci-Hz bands, and to hecto-Hz band.  
We expect our findings will motivate further investigation on prospects 
for multiband observation of \add{the} GW190521-like BBH systems, 
\rep{in addition to}{as much alike} the prototypical GW150914-like BBH system\add{s}~\cite{Abbott:2016blz}. 

Nevertheless, we emphasize that our study \rep{were}{was} performed with (very) simple methods. 
Therefore, our results should be only \textit{indicative} and \textit{tentative}. 
In the remainder of this section, 
we discuss what works remain to be done to refine our analysis in future, 
so that we can eventually make a strong scientific cases of multiband GW astronomy/physics.

\subsection{Assessment of prospects }

First, our methodology in Section~\ref{sec:method} should be replaced 
with more modern, sophisticated approaches to the GW data analysis; 
it will include (but not limited to), for example, 
the use of complete IMR
waveforms such as \rep{``effective-one-body'' (EOB)}{EOB} approach 
as well as \rep{``phenomenological'' (IMRPhenom)}{IMRPhenom} models 
(see, for examples, Ref.~\cite{Isoyama:2020lls} and references therein),  
full-fledged Bayesian posterior based techniques 
(see, for examples, Refs.~\cite{Veitch:2014wba,Biwer:2018osg,Ashton:2018jfp}), 
and more `realistic' noise and waveform models that account
for the sky-location of sources as well as orbital configurations 
of B-DECIGO~\cite{Kawamura:2020pcg} and LISA~\cite{Audley:2017drz}.


Second, our target GW190521-like BBH system was restricted 
to the nonprecessing, spinning, quasicircular configuration; 
although there is large uncertainty, the spin precession of GW190521 is estimated 
as nonzero~\cite{Abbott:2020tfl,Abbott:2020mjq}. 
Future studies are therefore needed to concern both the spin precession 
and orbital eccentricity of the BBH system: we will briefly discuss the eccentric, 
non-spinning BBH system in Appendix~\ref{app:ecc}. 
In general, the BH's spins in the precessing binary 
have not only their magnitudes but also orientations, 
which can be described by, for examples, 
the effective inspiral spin parameter $\chi_{\rm eff}$ 
(related to the components aligned with the orbital angular momentum),
and the precession spin parameter $\chi_{\rm p}$
(related to the components in the orbital plane for the inspiral GW waveform). 
Adding the orbital eccentricity to the (precessing) BBH system will be fully generic, 
and the waveform modeling becomes (much) more complicated. 
Despite that challenge, there is a considerable development 
of the analysis on fully generic binary systems~\cite{Ireland:2019tao}. 

Related with the point mentioned above, it should be noted that we have ignored 
subdominant $(\ell \neq 2,\,|m| \neq 2)$ harmonics in our GW waveform model.
They are more notable in the observed signal when system's mass ratio becomes smaller    
(such as the analysis of GW190814~\cite{Abbott:2020khf} 
with \add{the} mass ratio $q = 0.112^{+0.008}_{-0.009}$).
With the subdominant harmonics, one can access the source orientation 
to reduce uncertainty in the distance estimation~\cite{London:2017bcn}.


Third, we should note that there are two main hurdles
to analyze the ringdown GW signals; the low SNR with aLIGO, 
and the starting time of the ringdown phase that is 
\textit{a priori} unknown to the whole observed GW signal
(see, for examples, Refs.~\cite{Sakai:2017ckm,Carullo:2018sfu} 
for discussions on the starting time).
Our simple analysis is carried out with the single-mode waveform model~\eqref{eq:RDwave} 
as the template to analyse the ringdown phase, 
assuming that the starting time of the ringdown phase
to be $t = t_0$ which corresponds to the $f=f_{\rm R}$,
i.e., just after the end of the merger phase. 
While this choice does not affect our results of parameter estimation errors,
in practice, the best fit values (that should be obtained rather than assumed 
in the context of the full parameter estimation against the raw GW data) can be biased 
if one assumes the earlier starting time in the analysis
(see, for example, Figure~5 in Ref.~\cite{TheLIGOScientific:2016src}).
Although one may delay the starting time to avoid the bias 
in the parameter estimation, 
the SNR becomes much lower than the expected SNR 
with the assumption of $t = t_0$.

These obstacles will be overcome if one includes
higher overtones ($n > 0$) into
the ringdown GW analysis~\cite{Giesler:2019uxc},
for which a much larger SNR than the single-mode analysis will be expected. 
Indeed, a superposition of overtones ($n>0$)
in a single harmonic mode will be observed~\cite{Giesler:2019uxc,Isi:2019aib,Bhagwat:2019dtm,Ota:2019bzl} 
in the high SNR events (with ET). 
Another step-functional improvement of the ringdown 
GW analysis will be offered by using completely
different signal-analysis methods 
than the traditional matched filtering analysis, 
which may not be always optimal for the ringdown GW signal. 
There is ongoing work to assess the improvement 
due to new techniques for the ringdown GW signal analysis 
(such as Hilbert--Huang transformation, 
autoregressive modeling, and neural \rep{network}{networks})~\cite{Nakano:2018vay,Yamamoto:2020rse}.


Fourth, there are other \rep{types of GR test}{GR tests} that were 
not covered in Section~\ref{sec:implication}, 
but can be greatly improved using the multiband observation of GW195021-like BBH systems. 
Reference~\cite{Abbott:2020jks} performs various tests of GR with the BBH events in GWTC-2  
(see also the reviews by Carson and Yagi~\cite{Carson:2020rea} 
about the current and future test with GWs), including 
i) the IMR consistency test
between the inspiral and postinspiral phases divided at some cutoff frequency; 
ii) constraining deviations from GR
with parametric deformations to a predicted GR waveform model, 
iii)  ``BH spectroscopy''~\cite{Detweiler:1980gk,Dreyer:2003bv} with ringdown GWs which contains two (or possibly more)
QNMs~\cite{Berti:2016lat,Baibhav:2018rfk},
and so on.

The test i) is similar to our \rem{``}inspiral--ringdown\rep{''}{consistency} test 
directly using \rem{the} information of the merger phase, 
and makes the most of the GW waveform.
It has been pointed out that the values of this test 
with GW150914-like BBH systems will be maximized in the multiband observation~\cite{Carson:2019rda,Carson:2019kkh,Gnocchi:2019jzp}.
Our result suggests that the same will be true 
for \rem{the} GW190521-like ``intermediate-mass'' BBH system\add{s}, too.
Similarly, Gupta et al.~\cite{Gupta:2020lxa,Datta:2020vcj} have showed 
that the multiband observations of stellar- and intermediate-mass BBHs 
with LISA and 3G detectors will be only workable way
to carry out the most general version of test ii). 
Adding B-DECIGO (or any other planned GW detectors) in the deci-Hz band 
to the multiband analysis, 
we expect the precision of this test will be unprecedented.

Based on the generalized likelihood ratio test~\cite{Berti:2007zu}, the test iii) is performed
(see, for examples, Section~9.5 in Ref.~\cite{Akutsu:2020zlw} and references therein,
and also Ref.~\cite{Uchikata:2020wsp} for the future O5 era).
This test with aLIGO and Advanced Virgo alone is
quite challenging to have any conclusive result,
simply because of \rem{the too} low SNRs and \rem{the} larger parameter estimation errors in the ringdown phase.
In the 3G era, two (or possibly more) QNMs will be measurable~\cite{Berti:2016lat,Baibhav:2018rfk}.
It is also helpful to use the multiband observation
in order to optimize ground-based detectors 
via the forewarnings from the low-frequency, LISA band~\cite{Tso:2018pdv}. 
In either cases, \rep{multibanding}{the multiband observation} with B-DECIGO and ET will
give them additional advantage of 
being able to perform the best test of Kerr hypothesis
of remnant BHs via BH spectroscopy.


%

%

\vspace{6pt} 



\authorcontributions{
The authors contribute equally to this paper.
}

\funding{
H.~N. acknowledges support from JSPS KAKENHI Grant No. JP16K05347.
S.~I. acknowledges support from STFC through Grant No. ST/R00045X/1. 
S.~I. also thanks to networking support by the GWverse COST Action CA16104, 
“Black holes, gravitational waves and fundamental physics."
N.~S. and H.~N. acknowledge support from JSPS KAKENHI Grant No. JP17H06358.
}

\acknowledgments{
We would like to thank Carlos~O.~Lousto, James~Healy and 
Leor Barack for useful discussion.
\add{We also express our sincere gratitude to anonymous referees, 
who gave us valuable comments and kindly pointed out many typos in a previous version of this manuscript.} 
All the analytical and numerical calculations
in this paper have been performed with \textit{Maple} 
and Black Hole Perturbation Club (B.H.P.C.) codes~\cite{BHPC}.}

\conflictsofinterest{The authors declare no conflict of interest.} 



\appendixtitles{yes} 
\appendix

\section{Signal-to-noise ratio of GW190521-like eccentric BBH systems}
\label{app:ecc}

Throughout the bulk of this paper, we have looked at \add{the} GW190521-like BBH system 
under the assumption of a quasicircular BBH merger. 
While the quasicircular evolution of GW190521 is totally consistent 
with the LIGO/Virgo observation~\cite{Abbott:2020tfl,Abbott:2020mjq}, 
due to the lack of the inspiral GW signal long enough,
it appears that alternative scenarios, for example, 
GW190521 as an eccentric BBH merger also become\rem{s} relevant.
Indeed, Gayathri et al.~\cite{Gayathri:2020coq} demonstrated that 
observed GW190521 data could be explained as an equal-mass, 
highly-eccentric ($e=0.7$) BBH system 
\footnote{The orbital eccentricity ($e=0.7$) is provided as initial data 
for NR simulations at a frequency of 10\,Hz for a system 
with the total mass of $50\,M_{\odot}$
that is the orbital separation $\sim 24.5\, m_t$ 
for BBHs with the total mass $m_t$~\cite{Carlos_private}.
The NR waveforms are scaled by the total mass $m_t$.
}, 
and the estimated source parameters are quite different from
those derived from the quasicircular BBH scenario (recall Table~\ref{table:GW190521}): 
the primary mass $m_1^{\rm r}=102^{+7}_{-11}\,M_{\odot}$, 
the secondary mass $m_2^{\rm r}=102^{+7}_{-11}\,M_{\odot}$,
and the total mass $m_t^{\rm r}=204^{+14}_{-33}\,M_{\odot}$
in the source's rest frame,
the mass ratio $q=1$, the luminosity distance $D_L=1.84^{+1.07}_{-0.054}\,$Gpc
and the redshift $z=0.35^{+0.16}_{-0.09}$
(see Ref.~\cite{Gayathri:2020coq} for the spin parameters).
The possibility of GW190521 with nonvanishing eccentricity is also 
pointed out by Refs.~\cite{Romero-Shaw:2020thy,CalderonBustillo:2020odh}.

Like the quasicircular case, the multiband observation 
of eccentric GW190521-like BBH systems 
will once again help in distinguishing these two scenarios.  
Assuming a quadrupole GW generation from a Newtonian 
Kepler orbit~\cite{Peters:1963ux,Peters:1964zz}, 
the typical coalescing time $t_c^{\rm ecc}$ and 
the characteristic strain amplitude $h_{c,\,n}^{\rm ecc}$ 
of the $n$-th harmonic are 
(see, for example, Section~4.1 of Maggiore's text~\cite{Maggiore:1900zz} 
as well as Ref.~\cite{Amaro-Seoane:2018gbb}) 
\begin{equation}
 \label{eq:thc_ecc}
 t_c^{\rm ecc} \sim t_c^{\rm circ}\,(1 - e_0^2)^{7/2}\,,
\quad
 h_c^{\rm ecc} \sim h_c^{\rm circ}\,g(n,\,e) \,,
\end{equation}
where $t_c^{\rm circ}$ and $h_c^{\rm circ}$ are corresponding circular-inspiral results 
given in Eqs.~\eqref{eq:tc} and~\eqref{eq:hc}, respectively, 
and the function $g(n,\, e)$ will be defined momentarily. 
We see that the early \rep{inspial}{inspiral} phase of the eccentric binary is well within  
the B-DECIGO and LISA bands, too. 
Specifically, Holgado et al.~\cite{Holgado:2020imj} 
(see also the \rep{series of work}{works} 
by Amaro-Saoane~\cite{AmaroSeoane:2009yr,AmaroSeoane:2009ui,Amaro-Seoane:2018gbb}) 
pointed out that having deci-Hz GW observatories such as B-\rep{DEICGO}{DECIGO}, MAGIS~\cite{Graham:2017lmg} 
and TianGO~\cite{Kuns:2019upi} will be a key element 
to observe the eccentric inspiral GW signals in \rem{the} multiband \add{networks} 
because the GW signals may skip the LISA band entirely;  
$h_{c,\,n}^{\rm ecc}$ can be suppressed by a function of $g(n,\,e)$ significantly, 
depending on its eccentricity in the LISA band.

To better understand the visibility of eccentric BBH systems 
with aLIGO, ET, B-DECIGO and LISA,   
let us estimate the SNR of non-spinning, eccentric BBH 
inspirals accumulated in each band; 
see also Ref.~\cite{Loutrel:2020kmm}
for a recent review 
about waveform families for the eccentric binary systems.
\add{We also see various active works on eccentric waveform approximants~\cite{Nagar:2021gss,Arredondo:2021rdt,Setyawati:2021gom,Islam:2021mha}.}
The squared SNR averaged over the all-sky positions and binary orientations 
may be written as~\cite{Finn:2000sy,Dalal:2006qt,Moore:2014lga,Sesana:2016ljz} 
\footnote{
The expression for the average SNR~\eqref{eq:rho-ecc} recovers Eq.~\eqref{rho-ave} 
(of the inspiral part) in the circular orbit limit, $e \to 0$. 
Note that the harmonics are restricted to only $n = 2$ in the circular limit 
because one has $\lim_{e \to 0} g(n,\,e) = \delta_{2,\,n}$.
}
\begin{equation}
\label{eq:rho-ecc}
\rho^2_{\rm ave}
\approx
\frac{1}{5}\,\sum_{n}
\int_{f_{\mathrm {in}}}^{f_{\mathrm {end}}} 
\frac{h_{c,\,n}^2}{f_n\,S(f)} \, d\,(\ln f_n) \,,
\end{equation}
where $S(f)$ is the noise PSD for a given GW detector, and 
\begin{equation}
f_n 
=
n\,f_{\rm orb}
=
n\,f_{\rm orb}^{\rm r}\,(1 + z)^{-1}\,, 
\end{equation}
is the frequency of the harmonic in the observer frame, defined 
by the source's rest-frame frequency $f_{\rm orb}^{\rm r}$ of the Kepler orbit 
with the redshift $z$. 
The dimensionless characteristic strain $h_{c,\,n}$ 
of the $n$-th harmonic is~\cite{Finn:2000sy} 
\begin{equation}
\label{eq:hc_ecc}
h_{c,\,n}
\equiv
\frac{1 + z}{\pi\, D_{L}} \,
\sqrt{2 \frac{dE_n^{\rm r}}{d f_n^{\rm r}}} \,,
\end{equation}
where $D_L$ is the luminosity distance. 
Assuming the quadrupole formula applied to the Kepler orbit 
(with the chirp mass ${\cal M}$ in the observer frame)~\cite{Peters:1963ux,Peters:1964zz}, 
\begin{equation}
\label{dEdf-ecc}
\frac{dE_n^{\rm r}}{d f_n^{\rm r}}
\equiv
\frac{\pi}{3} 
\frac{{\cal M}^{5/3}}{(1+z)^2}\,
\frac{g(n,\,e)}{F(e)}\, \left( \frac{2}{n} \right)^{2/3}\,(\pi\, f_n)^{-1/3}\,,
\end{equation}
is the emitted GW energy per unit frequency $f_n^{\rm r}$ at the $n$-th harmonic 
measured in the source's rest frame, 
and we define 
\begin{align}
g(n,\,e) \equiv 
\frac{n^4}{32} \biggl\{ &
\left[
J_{n-2}(ne) - 2 e\,J_{n-1}(ne) + \frac{2}{n}J_n(ne)
+ 2 e\,J_{n+1}(ne) - J_{n+2}(ne)
\right]^2  \cr 
& + 
(1 - e^2)\, \left[
J_{n-2}(ne) - 2 \,J_{n}(ne) + J_{n+2}(ne)
\right]^2
+ \frac{4}{3 n^2} J^2_{n} (ne)
\biggr\}\,, \\
F(e) &\equiv 
\frac{1 + \frac{73}{24}\,e^2 + \frac{37}{96}\,e^4}{(1 - e^2)^{7/2}}
\left( = \sum_{n=1}^{\infty} g(n,\,e) \right)\,,
\end{align}
with the Bessel functions of the first kind $J_n(x)$
($n$: integer); 
see, for example, Ref.~\cite{Huerta:2015pva} for the derivation of Eq.~\eqref{dEdf-ecc}. 

For an inspiraling eccentric BBH, the evaluation of SNR through Eq.~\eqref{eq:rho-ecc} 
requires the knowledge of slowly-evolving orbital eccentricity $e$ and the frequency $f_n$ in time, 
under the gravitational radiation losses. Again, in the quadrupole formalism, 
it is given by~\cite{Peters:1963ux,Peters:1964zz}
\begin{equation}
\label{eq:f-e}
\frac{f_{\rm orb}}{f_{{\rm orb},\,0}}
=
\left[
\frac{1 - e_0^2}{1 - e^2}
\left(\frac{e}{e_0}\right)^{12/19}\,
\left(
\frac{1 + \frac{121}{304}e^2}{1 + \frac{121}{304}e_0^2}
\right)^{870/2299}
\right]^{-3/2} \,,
\end{equation}
with reference eccentricity $e_0$ 
and orbital frequency $f_{{\rm orb},\,0}$ (in the observer frame). 
One can set these constants by the values at the last stable orbit \add{(LSO)}
of the eccentric geodesic (of a test particle) 
in the \rep{Schcwarzschild}{Schwarzschild} geometry~\cite{Cutler:1994pb}: 
namely,  
$e_0 = e_{\rm LSO}$ and 
\begin{align}
\label{eq:LSO-ecc}
f_{{\rm orb},\,0} = f_{\rm LSO} 
& =
\frac{1}{2 \pi (1 + z) m_t^{\rm r}}
\left(
\frac{1 - e^2_{\rm LSO}}{6 + 2 e_{\rm LSO}}
\right)^{3/2}
\cr
& \sim  1.6 \times 10^{2}\,(1+z)^{-1} \left(\frac{m_t^{\rm r}}{204M_{\odot}}\right)^{-1}
\left(
\frac{1 - e^2_{\rm LSO}}{6 + 2 e_{\rm LSO}}
\right)^{3/2}\,{\rm Hz}
\,,
\end{align}
with the total mass $m_t^{\rm r}$ in the source's rest frame.
The frequency evolution~\eqref{eq:f-e} is therefore 
completely determined 
with a given single parameter $e_{\rm LSO}$.

We compute the averaged SNR $\rho_{\rm ave}$ given in Eq.~\eqref{eq:rho-ecc} 
for the GW190521-like, non-spinning, eccentric BBH\rem{s} with the source-frame masses 
$(m_1^{\rm r} ,\,m_2^{\rm r}) = (102\,M_\odot,\, 102\,M_\odot)$  
and the luminosity distance $D_L = 1.9\,{\rm Gpc}$ (i.e., the redshift $z \sim 0.35$), 
which mimics the eccentric BBH merger obtained 
by the NR simulations in Ref.~\cite{Gayathri:2020coq}. 
We assume the five year observation prior to the final merger 
determined by Eqs.~\eqref{eq:thc_ecc} and ~\eqref{eq:LSO-ecc}, 
and apply the same setup described in Section~\ref{subsec:num-setup} 
to each $n$-th harmonic of GW strains. 
Because the frequency evolution~\eqref{eq:f-e} is expressed 
in term of eccentricity, in practice, we change the integration variable 
of Eq.~\eqref{eq:rho-ecc} from $f_n$ to $e$ for the computational efficiency, 
making use of $d f_n = n |{df_{\rm orb}}/{de}|\, de$ with 
(see, for example, Ref.~\cite{Gondan:2017hbp}) 
\begin{equation}
\frac{df_{\rm orb}}{de}
=
-\frac{18}{19}\,\frac{f_{\rm orb}}{e}
\frac{F(e)}{(1 - e^2)^{9/2}
\,\left( 
1 + \frac{121}{304}e^2
\right)}\,.
\end{equation}
Also, the infinite summation over the harmonics $n$ in Eq.~\eqref{eq:rho-ecc} \rep{are}{is} truncated 
at some finite value of $n_{\rm max}$;  
we chose $n_{\rm max} = O(10^4)$ so that the resultant SNRs are 
guaranteed to have at least $3$ significant digits
\footnote{A good analytical estimator of $n_{\rm max}$ 
can be found in, for examples, Refs.~\cite{Drasco:2005kz,Mikoczi:2012qy}.}.

In Table~\ref{table:SNR_ecc}, we summarize the averaged SNRs $\rho_{\rm ave}$ 
of \add{the} GW190521-like, non-spinning, eccentric BBH\rem{s} accumulated in each GW band, 
for sample values of final eccentricities at the last stable orbit    
$e_{\rm LSO} = \{ 10^{-6},\,10^{-3},\,0.1,\,0.4,\,0.6 \}$. 
For references, the `initial' eccentricities at $f_{\rm in}$ for each band are also listed. 
It is approximated by solving Eq.~\eqref{eq:f-e} for $e$ 
at the detector's initial GW frequency of the second harmonics $f_{\rm orb} = f_{\rm in} / 2$. 
We found three main results:
i) B-DECIGO and ET always \rep{have the SNR larger than at least $10$ independent of the values of final eccentricity}{accumulate a total SNR greater than at least $10$, independent of the value of the final eccentricity} $e_{\rm LSO}$;  
ii) LISA has the detectable SNR only when $e_{\rm LSO} < 10^{-3}$. 
That is, the inspiral GW signal from \add{the} GW190521-like BBH 
that has \rep{a}{the} high-eccentricity in the aLIGO band 
would entirely skip the LISA band; 
iii) \rep{The}{the} SNR with B-DECIGO becomes bigger when $e_{\rm LSO}$ becomes smaller, 
while the SNR with ET and aLIGO shows the opposite behavior. 
Therefore, the \rep{mutibanding}{multiband observation} could provide much louder SNR 
than the single-band SNR across the full range of $e_{\rm LSO}$. 
Using Eq.~\eqref{Fisher-M} with the result in Table~\ref{table:SNR_ecc}, 
we find that the multiband SNR with B-DECIGO and ET are 
$\sim 180$
for $e_{\rm LSO} = \{10^{-6},\,10^{-3}\}$, 
$\sim 140$
for $e_{\rm LSO} = \{0.4,\,0.6\}$
and $\sim ~ 100$
for $e_{\rm LSO} = 0.1$.

\begin{table}[ht]
\caption{Averaged SNRs $\rho_{\rm ave}$ 
of the GW190521-like, non-spinning, eccentric BBHs  
accumulated in each band five years prior to coalescence, 
for \rem{a} given values of final eccentricity $e = e_{\rm LSO}$. 
We assume the source-frame component masses 
$(m_1^{\rm r} ,\,m_2^{\rm r}) = (102\,M_\odot,\, 102\,M_\odot)$, 
and the luminosity distance $D_L = 1.9\,{\rm Gpc}$ 
(i.e., the redshift $z \sim 0.35$)~\cite{Gayathri:2020coq}. 
The values in parentheses indicate the `initial' eccentricity 
estimated from the initial GW frequency ($f_{\rm in}$ defined in Section~\ref{sec:params-err}) 
of the second harmonic\rem{s} \add{mode} in each detectors. 
Note that aLIGO's `initial' eccentricity 
when $e_{\rm LSO} = 0.6$ is not displayed 
because the second harmonic\rem{s} \add{mode} is not detectable in this case. 
}
\centering
\begingroup
\renewcommand{\arraystretch}{1.2} 
\scalebox{0.8}[0.8]{
\begin{tabular}{l|ccccc}
\toprule
\quad & \multicolumn{5}{c}{SNR and eccentricity at $f_{\rm in}$} \\
\quad & 
\quad $e_{{\rm LSO}} = 10^{-6}$ &
\quad $e_{{\rm LSO}} = 10^{-3}$ &
\quad $e_{{\rm LSO}} = 0.1$ & 
\quad $e_{{\rm LSO}} = 0.4$ & 
\quad $e_{{\rm LSO}} = 0.6$  \\
\midrule
aLIGO \quad & 
\quad $4.95$ & 
\quad $4.95$ & 
\quad $5.55$ & 
\quad $1.07 \times 10^{1}$ & 
\quad $1.11 \times 10^{1}$ \\ 
\quad & 
\quad $(1.64 \times 10^{-6})$ & 
\quad $(1.64 \times 10^{-3})$ & 
\quad $(0.150)$ & 
\quad $(0.405)$ & 
\quad $(\cdots)$ \\ 
\midrule
ET \quad & 
\quad $7.64 \times 10^{1}$ & 
\quad $7.63 \times 10^{1}$ & 
\quad $8.73 \times 10^{1}$ & 
\quad $1.42 \times 10^{2}$ & 
\quad $1.42 \times 10^{2}$ \\ 
\quad & 
\quad $(8.98 \times 10^{-6})$ & 
\quad $(8.97 \times 10^{-3})$ & 
\quad $(0.505)$ & 
\quad $(0.746)$ & 
\quad $(0.793)$ \\ 
\midrule
B-DECIGO \quad & 
\quad $1.67 \times 10^{2}$ & 
\quad $1.70 \times 10^{2}$ & 
\quad $5.08 \times 10^{1}$ & 
\quad $1.85 \times 10^{1}$ & 
\quad $1.20 \times 10^{1}$ \\ 
\quad & 
\quad $(2.41 \times 10^{-3})$ & 
\quad $(0.709)$ & 
\quad $(0.982)$ & 
\quad $(0.992)$ & 
\quad $(0.993)$ \\ 
\midrule
LISA \quad & 
\quad $7.60$ & 
\quad $2.05$ & 
\quad $< 1.00$ & 
\quad $< 1.00$ & 
\quad $< 1.00$  
\\ 
\quad & 
\quad $(3.86 \times 10^{-3})$ & 
\quad $(0.985)$ & 
\quad $(> 0.999)$ & 
\quad $(> 0.999)$ & 
\quad $(> 0.999)$ \\ 
\bottomrule
\end{tabular}
}
\endgroup
\label{table:SNR_ecc}
\end{table}

Finally, although the parameter estimation is not the main focus here, 
we briefly discuss the potential accuracy of the eccentricity measurement. 
In the small-eccentricity and high-SNR limit, 
the orbital eccentricity may be measured 
within the fractional error\rem{s}~\cite{Nishizawa:2016jji} (see also Ref.~\cite{Seto:2016wom}) 
\begin{equation}
\delta {\hat e}_0 \sim 5\times 10^{-5} \,\frac{(1+z)^{5/3}}{\sqrt{2+3\alpha}} 
\left(\frac{{\cal M}^{\rm r}}{65.1M_\odot}\right)^{5/3}
 \left(\frac{f_0}{0.1\,{\rm Hz}}\right)^{5/3} 
\left( \frac{e_0}{0.1}\right)^{-1} \,,
\end{equation}
from Eq.~(2) in Ref.~\cite{Nishizawa:2016jji}
by using a rough approximation
with the quasicircular amplitude and the eccentric phase,
where $\delta {\hat e}_0$ denotes the parameter estimation error of $e_0$ (at the GW frequency $f_0$ for $n=2$)
normalized by the SNR.
Here, the power $\alpha$ is 
due to the approximation \add{of the} noise PSD by \rep{a}{the} power law,
$S_n \sim f^{2 \alpha}$ (assuming $\alpha>-2/3$), 
and B-DECIGO may have $\alpha = 1$, for example.  
This estimator implies that B-DECIGO would be able to precisely 
measure the eccentricity of GW190521-like BBH systems. 
Because we find that the `B-DECIGO + ET' combination always provides 
the multiband SNRs larger than $\sim 100$ independent of the value of $e_{\rm LSO}$, 
one might expect that this \rep{multibanding}{combination} best observes GW190521-like BBHs 
over the full range of eccentricity, 
helping to understand the population properties of BBH mergers~\cite{Abbott:2020gyp}.  
We will explore this possibility in future work.

\section{Some more noise power spectral densities 
of (next-generation) GW detectors}
\label{app:various_sensitivity}

In this appendix, we summarize some (fitting) curves of the noise PSD
\rep{of}{for} both ground and space-based, current and future GW detectors. 
These sensitivity curves are not used in the bulk of this paper, 
but it will serve as a convenient all-in-one-place summary with our notation; 
these curves with \rep{$f/(1\,{\rm Hz})$}{$f$ in unit of Hz} are shown in Figure~4 of Ref.~\cite{Kinugawa:2020tbg} 
(except DECIGO and TianQin).

\begin{itemize}
\item
``LIGO O3a-Livingston'' rough fitting curve
(during the first half of LIGO/Virgo third observing run by using Ref.~\cite{Aasi:2013wya}):
\begin{align}
S_n^{\rm O3a-L} = 
 \biggl(& { 2.13068\times 10^{-12}}\,{f}^{- 7.938724592}
 +{ 4.0\times 10^{-22}}\,{f}^{- 1.0}
 \cr & +{ 3.0\times 10^{-24}}+{ 1.74546
\times 10^{-27}}\,{f}^{ 1.178746922} \biggr)^{2}~~{\rm Hz}^{-1} \,.
\end{align}

\item
``LIGO O5'' rough fitting curve
(will be in the fifth observing run by using Ref.~\cite{Aasi:2013wya}):
\begin{align}
S_n^{\rm O5} = 
 \biggl( & 480985000.0\,{f}^{- 30.28419138} 
 +{ 6.63263\times 10^{-20}}\,{f}^{- 3.122716032} \cr 
& +
{ 6.15101\times 10^{-21}}\,{f}^{- 2.089976737}
+  { 1.32853\times 10^{-27}}\,{f}^{ 1.059219544}
 \biggr)^{2}~~{\rm Hz}^{-1} \,.
\end{align}

\item
``ET-B'' (another sensitivity curve for ET in Ref.~\cite{Hild:2008ng} 
other than Eq.~\eqref{noise-ET?}; 
see also, for examples, Ref.~\cite{Mishra:2010tp} and ET sensitivity page~\cite{ET-PSD}):
\begin{align}
S_n^{\rm ET-B} = { 1.0\times 10^{-50}}\, \biggl(&  45540.5\,{f}^{- 15.64}+ 6804.96\,
{f}^{- 2.145}
\cr &
+ 3.05853\,{f}^{- 0.12}+ 0.00258062\,{f}^{ 1.1} \biggr)^{2}~~{\rm Hz}^{-1} \,.
\end{align}

\item
``CE2'' rough fitting curve (for Cosmic Explorer presented in Ref.~\cite{Reitze:2019iox}):
\begin{align}
S_n^{\rm CE2} = 
\left( { 1.74408\times 10^{-16}}\,{f}^{- 8.908164528}+{ 2.0
\times 10^{-25}}+{ 8.23008\times 10^{-32}}\,{f}^{ 2.095903274} \right)^2~~{\rm Hz}^{-1} \,.
\end{align}

\item
``DECIGO'' (the noise PSD of the L-shaped configuration~\cite{Yagi:2011wg}):
\begin{align}
S_n^{\rm DECIGO} = 
\biggl\{ & 7.05 \times 10^{-48}
\left[
1 + \left( \frac{f}{f_p} \right)^2
\right]
+
4.8 \times 10^{-51}
f^{-4}
\left[
1 + \left( \frac{f}{f_p} \right)^2
\right]^{-1}\cr 
& \quad + 
5.53 \times 10^{-52}
f^{-4}\biggr\}~~{\rm Hz}^{-1} \,.
\end{align}
with $f_p \equiv 7.36$. 
Note that this expression accounted for the factor of $(\sqrt{3}/2)^{-2}$ 
due to DECIGO having arms that make an angle of $60^{\circ}$. 
\item
``TianQin''~\cite{Luo:2015ght,Hu:2018yqb} 
(the sky averaged noise PSD; the expression below is quoted 
from Eqs.~(9) and (10) of Ref.~\cite{Shi:2019hqa}):
\begin{align}
S_n^{\rm TianQin} = 
 { 3.0\times 10^{-51}}\, \biggl(&  
 0.009505539123\, {f^{-5}} + 95.05539123\,{f^{-4}}  + 0.07550033531\, {f^{-3}}  \cr 
& + 755.0033531\, {f^{-2}} + 3.703703703 \times 10^{10}
\cr &
+ 2.941767614 \times 10^{11}\,f^{2}
 \biggr)~~{\rm Hz}^{-1} \,.
\end{align}

\end{itemize}

\reftitle{References}





\end{document}